\documentclass[pre,twocolumn,amsmath,showpacs]{revtex4}

\usepackage[pdftex]{graphicx}
\usepackage{bm}
\usepackage{amssymb}
\usepackage{mathrsfs}
\usepackage{trfsigns}

\begin{document}
\title{Fractional Langevin Equation: Over-Damped, Under-Damped and Critical Behaviors.}

\author{S. Burov, E. Barkai}
\affiliation{Department of Physics, Bar Ilan University, Ramat-Gan 52900 Israel}

\begin{abstract}

The dynamical phase diagram of the fractional Langevin equation is investigated for harmonically bound particle. It is shown that critical exponents mark dynamical transitions in the behavior of the system. Four different critical exponents are found. (i) $\alpha_c=0.402\pm 0.002$ marks a transition to a non-monotonic under-damped phase, (ii) $\alpha_R=0.441...$ marks a transition to a resonance phase when an external oscillating field drives the system, (iii) $\alpha_{\chi_1}=0.527...$ and (iv) $\alpha_{\chi_2}=0.707...$ marks transition to a double peak phase of the ``loss" when such an oscillating field present. As a physical explanation we present a cage effect, where the medium induces an elastic type of friction. Phase diagrams describing over-damped, under-damped regimes, motion and resonances, show behaviors different from normal.

\end{abstract}
\maketitle

\date{\today}

\section{introduction}
In this paper we investigate the phenomenological description of stochastic
processes using a Fractional Langevin Equation (FLE) \cite{Adelman,Lutz,Pottier,Chaud,Hangii,Kop,Goychuk3}. While in simple systems the memory friction kernel is
an exponentially decaying function or a delta function, in complex
out of equilibrium systems the picture is in some cases different. Namely the relaxation is of a power
law type, and the particle may exhibit anomalous diffusion and relaxation \cite{Metzler1}.
Mathematically such systems are modeled using fractional calculus, e.g. $\frac{d^{1/2}}{dt^{1/2}}$. An example is the recent experiment on protein dynamics of the group of Xie~\cite{Xie3,Xie1}. There anomalous dynamics of the coordinate $x$, describing donor-acceptor distance was recorded, and a FLE (see Sec.~\ref{the_model}) was found to describe the experimental data. The motion of $x$ is bounded by a harmonic force field, and the equation of motion for the average $\langle x\rangle$ is
\begin{equation}
\ddot{\langle x\rangle}+\omega^2\langle x\rangle +
\gamma\frac{d^{\alpha}\langle x\rangle}{dt^{\alpha}}=0
\label{bar_averag1}
\end{equation}
where $0<\alpha<1$, $\omega$ is the harmonic
frequency and $\gamma>0$. For the case $\alpha=1$ we get the usual
damped oscillator~\cite{Kibble}. For such a normal case two types of
behaviors: the under-damped and the over-damped motions are found.
In the under-damped case $\langle x\rangle$ is oscillating, and
crossing the zero line, while for over-damped case $\langle
x\rangle$ is monotonically decaying with no zero crossing. For
$\alpha=1$, there exist a critical frequency $\omega_c=\frac{\gamma}{2}$
which separates these two types of motion. Here we explore a similar
scenario for the fractional oscillator, and find rich types of
physical behaviors. It is known~\cite{Argentina} that in the long
time limit all solutions $\langle x\rangle$ (i.e. any $0<\alpha$,
$0<\gamma$ and $0<\omega$ ) decay monotonically, some what like the
over-damped behavior of the usual oscillator, however now the decay
is of power law type. The interesting Physics occurs at shorter
times where the solution may exhibit different types of relaxation
and oscillations. We find that for $\alpha<\alpha_c$ the solution is
non-monotonic for any set of parameters ($\gamma,\omega
>0$). Thus we find a critical $\alpha$ which marks a dynamical transition in the behavior of the system.

We also investigate the response of a system described by Eq.~(\ref{bar_averag1}) to an external oscillating force $F_0\cos(\Omega t)$.
For the regular case of $\alpha=1$, a resonance is present if the frequency $\omega$ is larger than the critical value $\gamma/\sqrt{2}$. The behavior for $0<\alpha<1$ is quite different and we discover that the transition between $\alpha\rightarrow 1$ and $\alpha\rightarrow 0$ is not smooth. In particular we find another critical exponent $\alpha_R$, where for $\alpha<\alpha_R$ a resonance is always present. Other critical exponents are found for the imaginary part of complex susceptibility.
Our goal is to clarify the nature of solution to
Eq.~(\ref{bar_averag1}), investigate the meaning of fractional
critical frequency with and without external oscillating force and provide a mathematical tool box for finding
and plotting solutions of Eq.~(\ref{bar_averag1}). Our finding that critical $\alpha$s mark dynamical transitions is very surprising and couldn't be obtained without our mathematical treatment.

The paper is organized as follows. In Sec.~\ref{the_model} we present the FLE. In Sec.~\ref{GenSo} we present two different methods for solution for first order moments and examples are solved in Sec.~\ref{examples_1}. In Sec.~\ref{critical_w} we present different definitions for over-damped and under-damped motion and find the critical exponent $\alpha_c$. We also interpret our results from a more Physical point of view and discuss the cage effect as a viscoelastic property of the medium. In Sec.\ref{response_00} we introduce external oscillating force into the system and find the response for such force for the free (Sec.~\ref{Free_Motion}) and harmonically bounded (Sec.~\ref{Harmonic_b01}) particles. In Sec.~\ref{complex_01} we investigate the properties of the ``loss" - the imaginary part of complex susceptibility. A summary is provided in Sec.~\ref{Summary}, and the three Appendixes deal with some technical aspects.
A brief summary of some of our results was published~\cite{Burov}.

\section{The Model}
\label{the_model}
We consider the FLE 
\begin{equation}
m\frac{d^2x(t)}{dt^2}+\bar{\gamma}\int_0^t\frac{1}{(t-\grave{t})^{\alpha}}
\frac{dx}{d\grave{t}}\, d\grave{t}=F(x,t)+\xi(t) \label{int1}
\end{equation}
where $\bar{\gamma}>0$ is a generalized friction constant
($\gamma=\frac{1}{m}\bar{\gamma}\Gamma(1-\alpha)$), $0<\alpha<1$ is
the fractional exponent and $\xi(t)$ is a stationary, Fractional
Gaussian noise~\cite{Mandelb,Xie2} satisfying the
fluctuation-dissipation relation \cite{Kubo1}
\begin{equation}
\langle\xi(t)\rangle=0,\qquad \langle\xi(t)\xi(\grave{t})\rangle =
k_bT\bar{\gamma}|t-\grave{t}|^{-\alpha}. \label{stationary}
\end{equation}
$F(x,t)$ is an external force. We follow experiment~\cite{Xie1} and assume $F(x,t) = -m\omega^2x$, later in Sec.~\ref{response_00} we will have $F(x,t) = -m\omega^2x+F_0\cos(\Omega t)$.
In Laplace Space it is easy to show, using the convolution theorem
\begin{equation}
\begin{array}{l}
\displaystyle
\hat{x}(s)=\frac{s+\frac{1}{m}\beta(s)}{s^2+\frac{1}{m}s\beta(s)+\omega^2}x_0+\frac{1}{s^2
+\frac{1}{m}s\beta(s)+\omega^2}v_0\\
\ \\
\displaystyle\qquad+\frac{1}{s^2+\frac{1}{m}s\beta(s)+\omega^2}\xi(s)
\end{array}
\label{recLap1}
\end{equation}
where $x_0$ and $v_0$ are initial conditions and
\begin{equation}
\beta(s)=\bar{\gamma}\Gamma(1-\alpha)s^{\alpha-1}.
\label{pow_memory}
\end{equation}
All along this work the variable in the parenthesis defines the
space we are working in (e.g. $\hat{x}(s)$ is the Laplace Transform
of $x(t)$). Eq.~(\ref{int1}) with $\alpha=\frac{1}{2}$ and
$F(x,t)=-m\omega^2x$, describes single protein dynamics
\cite{Xie1}. An experimentally measured quantity is the normalized
correlation function
\begin{equation}
C_x(t)=\frac{{\langle}x(t)x(0)\rangle}{{\langle}x(0)^2\rangle}.
\label{corrx}
\end{equation}
In what follows, thermal initial conditions are assumed $\langle\xi(t)x(0)\rangle=0$, $\langle x(0)^2\rangle=k_bT/m\omega^2$ and $\langle x(0)v(0)\rangle=0$. From Eq.~(\ref{recLap1}) we find
\begin{equation}
\hat{C}_x(s)=\frac{s+\gamma s^{\alpha-1}}{s^2+\gamma
s^\alpha+\omega^2}.
\label{fractionLap}
\end{equation}

It is easy to show that $C_x(t)$ satisfies the following fractional Eq.
\begin{equation}
\ddot{C}_x(t)+\omega^2C_x(t)+\gamma\frac{d^{\alpha}C_x(t)}{dt^\alpha}=0,
\label{fracGen2}
\end{equation}
with the initial conditions $C_x(0)=1$ and $\dot{C}_x(0)=0$, where the fractional derivative is defined in the Caputo sense ~\cite{Samko,Miller}
\begin{equation}
\frac{d^{\alpha}f(t)}{dt^\alpha}=\,_0D_t^{\alpha-1}\left(\frac{df(t)}{dt}\right)
\label{caputo1}
\end{equation}
and $_0D_t^{\alpha-1}$ is the Riemann-Liouville fractional operator~\cite{Samko,Miller}
\begin{equation}
_{0}D_t^{\alpha-1}f(t)=\frac{1}{\Gamma(1-\alpha)}
\int_0^t(t-\grave{t})^{-\alpha}f(\grave{t})\,d\grave{t}.
\label{rieman}
\end{equation}
Note that another way to write Eq.~(\ref{int1}) is
\begin{equation}
\ddot{x}+{\gamma}\frac{d^\alpha x}{dt^\alpha}+\omega^2x=\xi(t)
\label{fracGen}
\end{equation}
hence the name Fractional Langevin equation is justified. For the force free particle ($F(x,t)=0$ in Eq.~(\ref{int1}))  $\langle x^2\rangle\propto t^\alpha$ \cite{Masoliver1,Kop}, this is a sub-diffusive behavior since $0<\alpha<1$. 

The FLE in general and Eq.~(\ref{fracGen2}) in particular can be
derived from the Kac-Zwanzig model of a Brownian particle coupled
to a Harmonic bath~\cite{Zwanzig}. Eq.~(\ref{fracGen2}) can be derived
also from the Fractional Kramers Equation ~\cite{Barkai2}. The use
of fractional-differential equations like
Eqs.~(\ref{fracGen2},\ref{fracGen}) became quite common in recent
years ~\cite{Metzler1,Sokol}, especially in the context of anomalous
diffusion ~\cite{Barkai1,Metzler1,Barkai3}. Several other fractional
oscillator equations were considered in the
literature~\cite{Achar,Ryabov,Zaslav} and general solutions for
fractional-differential equations of the type Eq.~(\ref{fracGen2})
are studied in the mathematical literature
~\cite{Kilbas,Adding,Miller}. In the next section we present a practical
recipe, a tool box, for the solution of Eq.~(\ref{fracGen2}), and
show how to plot its solution. Our methods are general and can be
applied to other linear fractional differential equations. From a
more physical point of view the following questions are addressed in
the next sections: (i) When is $C_x(t)$ positive? (i.e. similar to
over-damped behavior, $\alpha=1$). (ii) When is $C_x(t)$
non-monotonic? (similar to under-damped motion for $\alpha=1$) (iii)
What is the analogue to the critical frequency
$\omega_c=\frac{\gamma}{2}$ found for the $\alpha=1$ case. (iv) Does a critical exponent $\alpha_c$ exist and if so, what is it's value?

\section{The General Solution}
\label{GenSo}

 In this section we will present a recipe for an analytical solution of Eq.~(\ref{fracGen2}).

The formula for $C_x(t)$ in Laplace space is Eq.~(\ref{fractionLap})
and if $\alpha$ was an integer then performing an
Inverse Laplace Transform would be simple 
using analysis of poles \cite{Lavr,Doech}, because then the denominator
of Eq.~(\ref{fractionLap}) is a polynomial.
We assume $\alpha$ is of the form $\alpha=\frac{p}{q}$, where $q>p>0$ are
integers and $\frac{p}{q}$ is irreducible (i.e not equal to some other $\frac{l}{n}$ where $l<p$ and $n<q$ are integers).

\subsection{Method A}
\label{method_a}
We rewrite Eq.~(\ref{fractionLap}) as
\begin{equation}
\displaystyle{\hat{C}_{x}(s)}= \frac{s+\gamma
s^{\frac{p}{q}-1}}{s^2+\gamma
s^{\frac{p}{q}}+\omega^2}\,\frac{\hat{Q}(s)}{\hat{Q}(s)}=
\frac{\left(s+\gamma s^{\frac{p}{q}-1}\right)
\hat{Q}(s)}{\hat{P}(s)} \label{comp2}
\end{equation}
where $\hat{P}(s)$ is a polynomial in $s$. According to a
mathematical theorem~\cite{Miller} we can always find $\hat{Q}(s)$
such that the denominator of Eq.~(\ref{comp2})
\begin{equation}
\hat{P}(s)=(s^2+\gamma s^{\frac{p}{q}}+\omega^2)\hat{Q}(s)
\label{eliadd1}
\end{equation}
is a regular polynomial in s. The polynomial $\hat{Q}(s)$ is called
the complementary polynomial. The task of finding $\hat{Q}(s)$ is
simple, $\hat{Q}(s)$ is a polynomial in $s^{\frac{1}{q}}$ of degree
$2q(q-1)$
\begin{equation}
\hat{Q}(s)=\sum_{m=0}^{2q(q-1)}B_{m}s^{\frac{m}{q}} \label{qpolynom}
\end{equation}
with $B_{2q(q-1)}=1$. The coefficients $B_{m}$ are found by equating
all the coefficients in the expansion of the product
$\hat{Q}(s)(s^2+\gamma s^{\frac{p}{q}}+\omega^2)$, with non-integer
powers of $s$, to zero. This produces a linear system of $2q(q-1)$
equations for $2q(q-1)$ variables, which in principle is a solvable
problem.

We also assume that all $2q$ zeros of $\hat{P}(s)$ are distinct. The
generalization to the case where the zeros are not
distinct will be treated later, such a behavior is related to a critical frequency of the system. We use the partial fraction expansion
\begin{equation}
\frac{1}{\hat{P}(s)}=\sum_{k=1}^{2q}\frac{A_k}{s-a_k} \label{comp3}
\end{equation}
where $a_k$ are the solutions of $\hat{P}(s)=0$ and $A_k$ are
constants defined as
\begin{equation}
A_k=\frac{1}{\displaystyle\frac{d\hat{P}(s)}{ds}\mid_{s=a_k}}.
\label{Acoeff}
\end{equation}
It can be shown ~\cite{Miller}
that
\begin{equation}
\frac{s^m}{\hat{P}(s)}=\sum_{k=1}^{2q}\frac{a_k^mA_k}{s-a_k}\ , \
m=0,1,\dots,2q-1. \label{comp4}
\end{equation}
The numerator of Eq.~(\ref{comp2}) is written using the expansion
\begin{equation}
\displaystyle{\hat{Q}(s)\left(s+\gamma s^{\frac{p}{q}-1}\right)=}
\displaystyle\sum_{m=0}^{2q-1}\sum_{j=0}^{q-1}\tilde{B}_{m\,j}s^{m-\frac{j}{q}}.
\label{qmult}
\end{equation}
Hence using Eqs.~(\ref{comp2},\ref{comp3},\ref{qmult})
\begin{equation}
\displaystyle \hat{C}_{x}(s)=\sum_{m=0}^{2q-1}\sum_{j=0}^{q-1}\sum_{k=1}^{2q}
\frac{a_{k}^{m}\tilde{B}_{m\,j}A_{k}}{s-a_k}s^{-\frac{j}{q}}\;\displaystyle.
\label{finalLap}
\end{equation}
So finally we reduced the problem of calculating the Inverse Laplace
Transform of Eq.~(\ref{fractionLap}) to performing Inverse Laplace
Transform for
\begin{equation}
\displaystyle\frac{1}{s-a_k}\;\Laplace\quad e^{a_kt}
 \label{type1}
\end{equation}
and
\begin{equation}
\displaystyle\frac{1}{s^{\frac{j}{q}}(s-a_k)}\;\Laplace\quad
\frac{e^{a_kt}}{\Gamma(\frac{j}{q}){a_k}^\frac{j}{q}}\gamma(\frac{j}{q},a_kt)
\label{type2sol}
\end{equation}
where $\gamma(\frac{j}{q},a_kt)$ is a tabulated Incomplete Gamma Function~\cite{Abram}. Using the series expansion for
$\gamma(a,x)$
\begin{equation*}
\gamma(a,x)=\Gamma(a)x^ae^{-x}\sum_{n=0}^{\infty}\frac{x^n}{\Gamma(a+n+1)}
\end{equation*}
we can write Eq.~(\ref{type2sol}) by the means of
generalized Mittag-Leffler function \cite{Erdelyi}
\begin{equation}
\frac{1}{s^\frac{j}{q}(s-a_k)}\;\Laplace\quad
t^\frac{j}{q}{E}_{1,1+\frac{j}{q}}(a_kt)\;. \label{mittagLap}
\end{equation}
The
Mittag-Leffler function satisfies
\begin{equation}
E_{\eta,\mu}(y)={\displaystyle\sum_{n=0}^\infty\frac{y^n}{\Gamma(\eta{n}+\mu)}}
\label{mittag1}
\end{equation}
with 
\begin{equation}
E_{\eta,\mu}(y)\sim\,-\frac{y^{-1}}{\Gamma(\mu-\eta)}\;\;\;y\rightarrow\infty.
\label{mittagAs}
\end{equation}
 Actually we have finished our task, we can now perform
an Inverse Laplace Transform of expressions like
Eq.~(\ref{fractionLap}) and even more general expressions which can be presented as a fraction of polynomials with fractional powers. To
summarize, we need to follow four steps (i) Calculate $\hat{Q}(s)$, 
which is equivalent to the diagonalisation of a matrix of size
$\left(2q(q-1)\right)^2$. (ii) Find the zeros of $\hat{P}(s)$,
$a_k$, and the coefficients of partial fractions expansion ,$A_k$ of
Eq.~(\ref{Acoeff}). (iii) Find the coefficients $\tilde{B}_{m j}$
for Eq.~(\ref{qmult}) and write $\hat{C}_x(s)$ as the sum in
Eq.~(\ref{finalLap}). (iv) Use
Eqs.~(\ref{type1},\ref{type2sol},\ref{mittagLap}) to inverse Laplace
transform Eq.~(\ref{finalLap}), we find
\begin{equation}
\displaystyle C_{x}(t)=\sum_{m=0}^{2q-1}\sum_{j=0}^{q-1}\sum_{k=1}^{2q}
\frac{a_{k}^{m-j/q}\tilde{B}_{m\,j}A_{k}e^{a_kt}}{\Gamma(\frac{j}{q})}
\gamma(\frac{j}{q},a_kt)
\label{finalsol1}
\end{equation}
where for $j=0$,  $\displaystyle{\frac{\gamma(\frac{j}{q},a_kt)}{\Gamma(\frac{j}{q})}=1}$, or using Eq.~(\ref{mittagLap})
\begin{equation}
\displaystyle C_{x}(t)=\sum_{m=0}^{2q-1}\sum_{j=0}^{q-1}\sum_{k=1}^{2q}
a_{k}^{m}\tilde{B}_{m\,j}A_{k}t^{\frac{j}{q}}E_{1,1+\frac{j}{q}}(a_kt)
\label{finalsol2}
\end{equation}
and for $j=0$ $E_{1,1+\frac{j}{q}}(a_kt)=e^{a_kt}$.
Now we have a practical tool
for finding an explicit analytical solution for the fractional damped harmonic oscillator. Since $\gamma(a,x)$ is tabulated in programs like Mathematica, the solution which is a finite sum of such incomplete gamma functions, can be plotted.
\begin{figure}
\begin{center}
\includegraphics[width=\columnwidth]{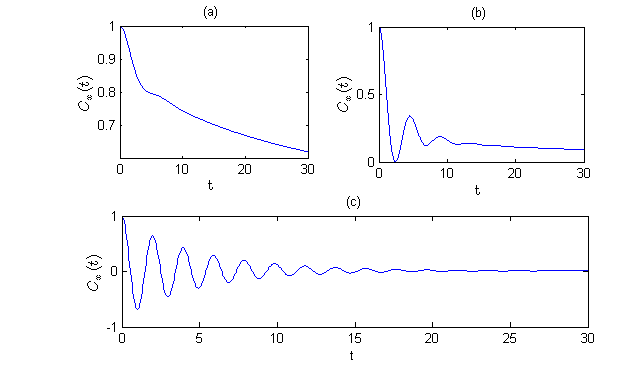}
\end{center}
\caption{ The short time behavior for $C_x(t)$ with
$\alpha=\frac{1}{2}$ and $\gamma=1$, versus $t$. Three types of
solutions are presented {\bf (a)} $\omega=0.3$ and the function decays
monotonically. {\bf (b)} $\omega=\omega_z\approx 1.053$ the transition between
motion with and without zero crossing, $C_x(t)=0$ at a single point
in time, and $C_x(t)$ does not cross the zero line. {\bf (c)} $\omega=3$  the
under-damped regime. Note that for large $t$, $C_x(t)>0$. }
\label{fig1}
\end{figure}

\subsection{Method B}
\label{method_b} The task of finding $\hat{Q}(s)$ is sometimes
difficult since as described in Sec.~\ref{method_a}, generally one
must solve a linear system of $2q(q-1)$ equations with $2q(q-1)$
variables. But for the special case of Eq.~(\ref{fractionLap}) we
will provide a simple method for finding $\hat{Q}(s)$ and
$\hat{P}(s)$. By writing
\begin{equation}
\begin{array}{l}
C_x(t)=\\
\displaystyle{
\frac{s+\gamma s^{\frac{p}{q}-1}}{(s^2+\omega^2)^q+(-1)^{q-1}\gamma^q s^p}
\left( \frac{\left(s^2+\omega^2\right)^q+(-1)^{q-1}\gamma^q
s^{p}}{s^2+\gamma s^{\frac{p}{q}}+\omega^2} \right) }
\end{array}
\label{metb_01}
\end{equation}
we can write
\begin{equation}
\hat{Q}(s)=\frac{\left(s^2+w^2\right)^q+(-1)^{q-1}\gamma^q
s^{p}}{s^2+\gamma s^{\frac{p}{q}}+w^2} \label{threeQuoter3}
\end{equation}
and
\begin{equation}
\hat{P}(s)=\left(s^2+\gamma s^{\frac{p}{q}}+w^2\right)\hat{Q}(s)
=\left(s^2+\omega^2\right)^q+(-1)^{q-1}\gamma^q s^p .
\label{threeQuoter4}
\end{equation}
One easily sees that indeed $\hat{P}(s)$ is a polynomial in $s$. As
for $\hat{Q}(s)$, it is found by standard method of division of two
polynomials in $s^{1/q}$ and it is also a polynomial
in $s^{1/q}$. Since the fraction of two polynomials
\begin{equation}
\frac{\left(y^{2q}+\omega^2\right)^q+(-1)^{q-1}\gamma^q{y^{qp}}}{y^{2q}+\gamma y^{p}+\omega^2}
\label{threeQuoter2}
\end{equation}
is also a polynomial in $y$, we see that any solution of $y^{2q}+\gamma
y^{p}+\omega^2=0$ is also a solution of
$\left(y^{2q}+w^2\right)^q+(-1)^{q-1}\gamma^q{y^{qp}}=0$, and by
performing a substitute $y=s^\frac{1}{q}$ into
Eq.~(\ref{threeQuoter2}) we find that indeed $\hat{Q}(s)$ is a
polynomial in $s^{1/q}$. After finding $\hat{Q}(s)$ and $\hat{P}(s)$ explicitly we can return to method A and use Eqs.~(\ref{comp3}-\ref{finalsol2}) in order to write down the final solution.

\begin{figure}
\begin{center}
\includegraphics[width=\columnwidth]{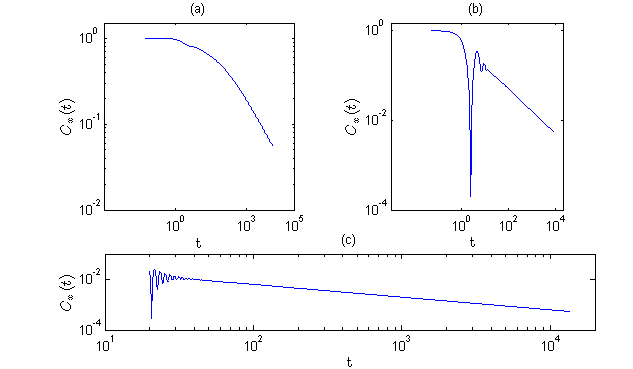}
\end{center}
\caption{ The long time behavior for $C_x(t)$ with
$\alpha=\frac{1}{2}$ and $\gamma=1$,versus $t$, on a loglog scale.
Three types of solutions are presented {\bf (a)} $\omega=0.3$ over-damped
limit, monotonic decay. {\bf (b)} $\omega=\omega_z\approx 1.053$ the
transition between motion with and without zero crossing. {\bf (c)}
$\omega=3$  the under damped regime, for large $t$, $C_x(t)>0$ (for short
time $C_x(t)<0$, not shown). Notice that non-monotonic decay in the
under-damped case is observed only for short times, while for
long-times $C_x(t)\propto t^{-\frac{1}{2}}$. } \label{fig4}
\end{figure}

\section{Examples $\alpha=\frac{1}{2}$ and $\alpha=\frac{3}{4}$.}
\label{examples_1}
\subsection{$\alpha=\frac{1}{2}$}
The $\alpha=\frac{1}{2}$ case was recently measured in experiment ~\cite{Xie1},
 and so it will be our first illustration for the method.
 From Eq.~(\ref{fractionLap}) we get
\begin{equation}
\hat{C}_x(s)=\frac{s+\gamma s^{-\frac{1}{2}}}{s^2 +\gamma
s^{\frac{1}{2}}+w^2}. \label{mexmp1}
\end{equation}
The first step is to find the complementary polynomial of the
denominator on the right hand-side of Eq.~(\ref{mexmp1}). Using
method A of the previous section one can easily see that the
coefficients of the complementary polynomial $\hat{Q}(s)$ are
\begin{equation*}
B_4=1,\,B_3=B_2=0,\,B_1=-\gamma,\,B_0=\omega^2
\end{equation*}
and
\begin{equation}
\hat{Q}(s)=s^2-\gamma s^{\frac{1}{2}}+\omega^2 \label{polexmp1}
\end{equation}
since
\begin{equation}
\hat{P}(s)=\hat{Q}(s)\left( s^{2}+\gamma
s^{\frac{1}{2}}+\omega^2\right)=s^4+2\omega^2s^2-\gamma^2s+\omega^4.
\label{halph_explain1}
\end{equation}
We rewrite Eq.~(\ref{mexmp1}) as
\begin{equation}
\hat{C}_x(s)= \frac{s^3+s\omega^2-\gamma^2+\gamma
\omega^2s^{-\frac{1}{2}}}{(s^2+\omega^2)^2-\gamma^2s} \label{mexmp2}
\end{equation}
and notice that the degree of the denominator of Eq.~(\ref{mexmp2}),
i.e. $\hat{P}(s)$, is $4$. Its zeros are easily found using Ferrari
formula ~\cite{Abram}, and we call them $a_k$, $k=1,\dots,4$. The
coefficients of the partial fraction expansion $A_k$ are given by
Eq.~({\ref{Acoeff})
\begin{equation}
A_k=\frac{1}{4a_k(a_k^2+\omega^2)-\gamma^2}. \label{eliadd2}
\end{equation}
The partial fraction expansion is found using Eq.~(\ref{finalLap})
\begin{equation}
\hat{C}_x(s)= \sum_{m=0}^{3}\sum_{j=0}^{1}\sum_{k=1}^{4}
\frac{a_{k}^{m}\tilde{B}_{m\,j}A_{k}}{s-a_k}s^{-\frac{j}{q}}
\label{halph_explain2}
\end{equation}
and the $\tilde{B}_{m\,j}$ in Eq.~(\ref{qmult}) are found using the numerator of Eq.~(\ref{mexmp2})
\begin{equation}
\tilde{B}_{3\,0}=1,\,\tilde{B}_{1\,0}=\omega^2,\,\tilde{B}_{0\,0}=-\gamma^2,\,
\tilde{B}_{0\,1}=\gamma \omega^2 \label{beegis}
\end{equation}
and other elements of the matrix $\tilde{B}_{m\,j}$ are equal 0.
Using Eq.~(\ref{finalsol1}) the solution is
\begin{widetext}
\begin{equation}
C_x(t)=\sum_{k=1}^4
{\left(\left(-\gamma^2+\omega^2a_k+a_k^3\right)A_ke^{a_kt}+\gamma \omega^2A_kt^{
\frac{1}{2}}E_{1,\frac{3}{2}}(a_kt)\right)}. \label{exmpsol1}
\end{equation}
\end{widetext}

\begin{figure}
\begin{center}
\includegraphics[width=\columnwidth]{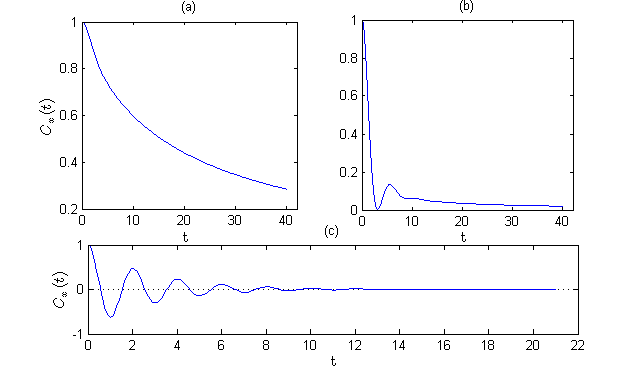}
\end{center}
\caption{ The short time behavior for $C_x(t)$ with
$\alpha=\frac{3}{4}$ and $\gamma=1$, versus $t$. Three types of
solutions are presented {\bf (a)} $\omega=0.3$ and the function decays
monotonically. {\bf (b)} $\omega=\omega_z\approx 0.965$ the transition between
motion with and without zero crossing, $C_x(t)$ does not cross the
zero line. {\bf (c)} $\omega=3$  the under-damped regime. } \label{fig3}
\end{figure}
By using a series expansion and some algebra we find $a_k$ and Eq.~(\ref{exmpsol1}) has the following asymptotic behavior
\begin{equation}
\displaystyle
C_x(t)= \left\{
\begin{array}{ll}
\displaystyle
1-\frac{1}{2}\omega^2t^2+\frac{\gamma \omega^2}{\Gamma(\frac{9}{2})}t^\frac{7}{2}
&
\;t\rightarrow0\\
\displaystyle \frac{\gamma}{\omega^2\Gamma(\frac{1}{2})}t^{-\frac{1}{2}}-
\left(\frac{\gamma}{w^2}\right)^3\,\frac{t^{-\frac{3}{2}}}{2\Gamma(\frac{1}{2})}
&  \;t\rightarrow\infty\quad.\\
\end{array}
\right\},
 \label{exmpas}
\end{equation}
We see that $C_x(t)$ for long times decays as a power-law,
which is the signature of slow relaxation and anomalous diffusion. The same asymptotic results are
found by applying Tauberian Theorems ~\cite{Feller} to
Eq.~(\ref{mexmp1}).

The asymptotic expansions Eq.~(\ref{exmpas}) provides the behavior
for long (and short) times, but the intermediate behavior is not
obvious. Using the exact solution Eq.~(\ref{exmpsol1}) we plot
$C_x(t)$ for various values of $\gamma$ and $\omega$ in Fig. \ref{fig1}
and Fig. \ref{fig4}. Three types of behaviors exist (i) Monotonic
decay of the solution - Fig. \ref{fig1}(a).(ii) Non-monotonic decay in the
non-negative half of the plane, $C_x(t)\geq 0$,  - Fig. \ref{fig1}(b). (iii)
Oscillations of the solution, where $C_x(t)$ also takes
negative values - Fig. \ref{fig1}(c). These are typical behaviors of
the solution which we found also in other parameter set (not shown).
From Fig.~\ref{fig1} we identify $\omega=\omega_z=1.053$ as a fractional
critical point, in the sense that if $\omega>\omega_z$ we have zero crossings
for $C_x(t)$. For $\alpha=\frac{1}{2}$ there exist also another
fractional critical point $\omega_m=0.426$ where for $\omega<\omega_m$ $C_x(t)$ is
monotonically decaying - Fig. \ref{fig1}(a). We will soon discuss
$\omega_z$ and $\omega_m$ more generally.

\subsection{$\alpha=\frac{3}{4}$}
In this subsection we demonstrate the solution for $\alpha=\frac{3}{4}$ using method B. Our goal is to invert Eq.~(\ref{fractionLap}).
\begin{equation}
\hat{C}_x(s)= \frac{s+\gamma s^{\frac{3}{4}-1}}{s^2+\gamma
s^\frac{3}{4}+w^2}\qquad. \label{threeQouter1}
\end{equation}
We find the complementary polynomial $\hat{Q}(s)$ and $\hat{P}(s)$,
using Eq.~(\ref{threeQuoter3}) and Eq.~(\ref{threeQuoter4})
\begin{equation}
\hat{P}(s)=\left(s^2+\omega^2\right)^4-\gamma^4 s^3
\label{threeQuoter5}
\end{equation}
and
\begin{figure}
\begin{center}
\includegraphics[width=\columnwidth]{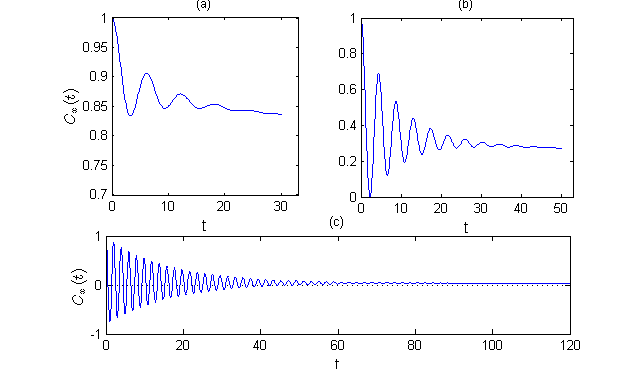}
\end{center}
\caption{ The short time behavior for $C_x(t)$ with
$\alpha=\frac{1}{5}$ and $\gamma=1$, versus $t$. The three types of
solutions are presented {\bf (a)} $\omega=0.3$ and the function
oscillates above zero. {\bf (b)} $\omega=\omega_z=1.035$ the transition
between motion with and without zero crossing, $C_x(t)$ does not
cross the zero line. {\bf (c)} $\omega=3$  oscillations with zero
crossing for short times, for long times $C_x(t)>0$ and no
oscillations are observed. For this case solution with monotonic
decay are not found since $\alpha=\frac{1}{5}<\alpha_c =0.402$. } \label{fig2}
\end{figure}
\begin{widetext}
\begin{equation}
\displaystyle{\hat{Q}(s)=\frac{\left(s^2+\omega^2\right)^4-\gamma^4 s^{3}}{s^2+\gamma s^{\frac{3}{4}}+\omega^2}=}
\displaystyle{s^6-\gamma s^\frac{19}{4}+3\omega^2s^4+\gamma^2s^\frac{7}{2}-2\omega^2\gamma s^\frac{11}{4}
-\gamma^3s^\frac{9}{4}+3\omega^4s^2+\omega^2\gamma^2s^\frac{3}{2}-\gamma \omega^4s^\frac{3}{4}+\omega^6}
\label{threeQuoter6}
\end{equation}
 and so we can write Eq.~(\ref{threeQouter1}) as
\begin{equation}
\hat{C}_x(s)=\frac{ s^7+3\omega^2s^5+\gamma
\omega^2s^\frac{15}{4}+3\omega^4s^3-\gamma^2\omega^2s^\frac{5}{2}-\gamma^4s^2
+2\gamma
\omega^4s^\frac{7}{4}+\gamma^3\omega^2s^\frac{5}{4}+\omega^6s-\gamma^2\omega^4s^\frac{1}{2}
+\gamma
\omega^6s^{-\frac{1}{4}}}{\left(s^2+\omega^2\right)^4-\gamma^4s^3}.
\label{threeQuoter7}
\end{equation}
\end{widetext}
The degree of the denominator of Eq.~(\ref{threeQuoter7}) is $8$, so
the zeros of the polynomial $\hat{P}(s)$ could only be found
numerically using a program like Mathematica. As in the previous
subsection we call the zeros of $\hat{P}(s)$  $a_k$ , $k=1,\dots,8$
and the coefficients of partial fraction expansion $A_k$ are found
using Eq.~(\ref{Acoeff})
\begin{equation}
A_k=\frac{1}{8a_k\left(a_k^2+\omega^2\right)^3-3\gamma^4a_k^2}.
\label{threeQuoter8}
\end{equation}
Writing the partial fraction expansion using Eq.~(\ref{finalLap})
\begin{equation}
\displaystyle \hat{C}_{x}(s)=\sum_{m=0}^{7}\sum_{j=0}^{3}\sum_{k=1}^{8}
\frac{a_{k}^{m}\tilde{B}_{m\,j}A_{k}}{s-a_k}s^{-\frac{j}{q}}\;\displaystyle,
\end{equation}
where the coefficients $\tilde{B}_{m\,j}$ are found using the numerator of Eq.~(\ref{threeQuoter7})
\begin{equation}
\tilde{B}=\left(\begin{array}{c c c c}
0&\gamma \omega^6&0&0 \\
w^6&0& -\gamma^2\omega^4&0 \\
-\gamma^4&2\gamma \omega^4&0&\gamma^3\omega^2 \\
3\omega^4&0&-\gamma^2\omega^2&0 \\
0&\gamma \omega^2&0&0 \\
3\omega^2&0&0&0 \\
0&0&0&0 \\
1 & 0 & 0& 0
\end{array}\right).
\label{threeQuoter9}
\end{equation}

Using Eq.~(\ref{finalsol1}) the solution is\newline
\begin{widetext}
\begin{equation}
\begin{array}{l}
\displaystyle{C_x(t)=}
\displaystyle{ \sum_{k=1}^{8} A_ke^{a_kt}\left(a_k\omega^6-a_k^2\gamma^4+3a_k^3\omega^4+3a_k^5\omega^2+a_k^7\right)} \\
\qquad \displaystyle{
+A_ke^{a_kt}\left(\frac{\gamma\gamma(\frac{1}{4},a_kt)}{\Gamma(\frac{1}{4})}
\left(\omega^6a_k^{-\frac{1}{4}} +2\omega^4a_k^\frac{7}{4}
+\omega^2a_k^\frac{15}{4} \right)
+\frac{\gamma^2\gamma(\frac{1}{2},a_kt)}{\Gamma(\frac{1}{2})} \left(
\omega^4a_k^\frac{1}{2}+\omega^2a_k^\frac{5}{2} \right)
+\frac{\omega^2a_k^\frac{5}{4}\gamma^3\gamma(\frac{3}{4},a_kt)}{\Gamma(\frac{3}{4})}
\right) }
\end{array}
\label{threeQuoter10}
\end{equation}
or using Eq.~(\ref{finalsol2})
\begin{equation}
\begin{array}{l}
\displaystyle{C_x(t)=}
\displaystyle{ \sum_{k=1}^{8} A_ke^{a_kt}\left(a_k\omega^6-a_k^2\gamma^4+3a_k^3\omega^4+3a_k^5\omega^2+a_k^7\right)} \\
\qquad \displaystyle{
+A_k\left(\gamma t^\frac{1}{4}E_{1,\frac{5}{4}}(a_kt) \left(\omega^6
+2\omega^4a_k^2 +\omega^2a_k^4 \right)
+\gamma^2t^\frac{1}{2}E_{1,\frac{3}{2}}(a_kt) \left( \omega^4a_k+\omega^2a_k^3
\right) +\omega^2a_k^2t^\frac{3}{4}\gamma^3E_{1,\frac{7}{4}}(a_kt)
\right). }
\end{array}
\label{threeQuoter11}
\end{equation}
\end{widetext}
The behavior described by Eq.~(\ref{threeQuoter11}) is plotted at Fig. \ref{fig3}. The three typical types of behavior are shown, which are similar to the behavior for the $\alpha=\frac{1}{2}$ case Fig.~\ref{fig1}. The values of critical points for $\alpha=\frac{3}{4}$ are $\omega_m\approx 0.707$ (and so for $\omega=0.3$ we observe in Fig.~\ref{fig3}(a) a monotonic decay) and $\omega_z\approx 0.965$, a case plotted in Fig.~\ref{fig3}(b).
Finally the case $\alpha=\frac{1}{5}$ is shown in Fig.~\ref{fig2}. The difference between $\alpha=\frac{1}{5}$ and the former cases is that for $\alpha=\frac{1}{5}$ $\omega_m=0$ and so we never observe a monotonic decay of the solution. More generally, this behavior is obtained for any $\alpha<\alpha_c\simeq 0.402$, as we shall soon show.

\section{Definition of over and under damped motion}
\label{critical_w}
As mentioned in the introduction, when dealing with the normal damped oscillator one gets two types of solutions - over-damped
and under-damped, the transition between these two behaviors happens at some point $\omega_c$ called the critical point.
For the over-damped motion $\langle x\rangle>0$ for any $t$ when $\langle x(t=0)\rangle>0$, and there are no oscillations, and for the under-damped case $\langle x(t)\rangle$ oscillates and crosses the zero line. For the fractional oscillator, we notice that there are different types of behaviors. From Figs.~(\ref{fig1},\ref{fig3},\ref{fig2}) one notices that for short times there is an oscillating behavior either with and without zero-crossings or a monotonic decay type of behavior. So as in the regular damped oscillator we need to define the transition between these behaviors.
We propose three definitions for the point of transition between
over-damped and under-damped motions, these are based on the various
definitions that exist for the regular damped oscillator and give
the same result for $\alpha=1$. The first option is to take the
frequency $\omega_c$ for which two solutions of $\hat{P}(s)=0$
coincide, i.e. an appearance of a pole of a second order for
$\hat{C}_x(s)$, and the general solution of Sec.~\ref{GenSo} must be modified, as explained in
Appendix~\ref{appa_a}. The second option is to take the minimal
frequency $\omega_z$ at which the solution $C_x(t)$ crosses the zero
line and the third is to take the minimal frequency $\omega_m$ at
which $\frac{dC_x(t)}{dt}$ crosses the zero line (i.e. $C_x(t)$ is
no longer a monotonically decaying function). For regular damped
oscillator $\omega_c=\omega_z=\omega_m$, but in fractional case this is
generally not true.

Another difference between the fractional oscillator and the regular one is the distinction between short and long time behavior when $0<\alpha<1$.
The asymptotic behavior for general $\alpha$ is obtained using general properties of polynomial solutions \cite{Miller} with Eq.~(\ref{mittag1}) and Eq.~(\ref{mittagAs}) or using the Tauberian theorem \cite{Feller}
\begin{equation}
\begin{array}{lr}
\displaystyle{C_x(t)}\approx1-\frac{1}{2}\omega^2t^2+
\frac{\omega^2\gamma}{\Gamma(5-\alpha )}t^{4-\alpha} & \qquad{t\to0}
\end{array}
\label{difin_1_0}
\end{equation}
\begin{equation}
\begin{array}{lr}
\displaystyle{C_x(t)}\approx
\frac{\gamma}{\omega^2\Gamma(1-\alpha)}t^{-\alpha}
 & \qquad{t\to\infty}
\end{array}
\label{difin_1}
\end{equation}
where the large $t$ expression was obtained in \cite{Xie2,Xie1,Argentina}. The applicability of Eq.~(\ref{difin_1}) is possible only under two conditions, the first one is obvious from Eq.~(\ref{difin_1}) and is
\begin{equation}
\left(\frac{\omega^2}{\gamma}\right)^{1/\alpha}t\gg 1,
\label{cond_app_01}
\end{equation}
while the second is
\begin{equation}
|a|t\gg 1,
\label{cond_app_02}
\end{equation}
where $|a|$ is an absolute value of a zero of the enumerator in Eq.~(\ref{fractionLap}) ($s^2+\gamma s^{\alpha}+\omega^2$), it also could be found from $\hat{P}(s)=0$. Mathematically the second condition is the magnitude of the radius of convergence for the power series expansion of $\hat{C}_x(s)$ near $s=0$ (Tauberian Theorem), and is given by the non-fictitious poles of $\hat{C}_x(s)$.
From Eq.~(\ref{difin_1}), $C_x(t)>0$ and for large $t$ decays as a power-law (see Fig.~\ref{fig4}).

From a more physical point of view, the FLE formalism was used~\cite{Xie1} to describe the fluctuation of the distance between a fluorescein-tyrosine pair within a single protein on a time scales of $1$ msec up to $10^2$ sec, and a power-law decay was observed (lately a theoretical model of Fractons was proposed in order to explain such phenomena~\cite{Granek}). Recent molecular dynamics simulations~\cite{Luo} studied the fluctuations of a donor-acceptor distance for a single protein, for short time scales $10^{-9}$ sec and oscillations of the autocorrelation function were observed. The scenario of oscillations for short-times and a power-law decay for long-times set's well with the description by our solutions of the  FLE.

\subsection{Critical point $\omega_c$}
\label{W_C} For the case $\alpha=\frac{1}{2}$ we have four solutions
for $\hat{P}(s)=0$ ($a_k$ $k=1,\dots,4$) which are plotted in Fig.
\ref{fig5}. At one point two solutions coincide, we call this point
the critical point at which $\omega=\omega_c$. For
$\omega=\omega_c$, the solution for $C_x(t)$ Eq.~(\ref{exmpsol1})
must be modified because our general method derived Sec~\ref{GenSo}
is not valid. For $\alpha=\frac{1}{2}$ and $\omega=\omega_c=\frac{1}{2^{2/3}}\sqrt{(3/2)(1/2)^{1/3}}\gamma^{2/3}$ (see Eq.~\ref{wc_7})
Eq.~(\ref{exmpsol1}) is (see Appendix, Eq.~(\ref{app11}))
\begin{widetext}
\begin{equation}
C_x(t)=\sum_{m=0}^{3}\sum_{j=0}^{1}\tilde{B}_{mj}t^{\frac{j}{2}}
\left[ \sum_{k=1}^{3}
a_k^{m}A_k
E_{1,1+\frac{j}{2}}(a_kt)+
a_{3}^{m}\tilde{A}
\left((t+ma_{3}^{-1})E_{1,1+\frac{j}{2}}(a_{3}t)
-\frac{j}{2}tE_{1,2+\frac{j}{2}}(a_{3}t)\right)
\right]
\label{wc_2}
\end{equation}
\end{widetext}
where $a_1\neq a_2\neq a_3$ and $a_4=a_3$ are zeros of $\hat{P}(s)$
given by Eq.~(\ref{halph_explain1}), $\tilde{B}_{mj}$ are found
using Eq.~(\ref{beegis}), $A_1$ and $A_2$ by Eq.~(\ref{eliadd2})
,$A_3=-(A_1+A_2)$ by Eq.~(\ref{app03}) and $\tilde{A}$ is defined in
Eq.~(\ref{app04}).

We emphasize that the critical point $\omega_c$ does not always
exist. In Fig.~\ref{fig6} we plot the $10$ solutions of
$\hat{P}(s)=0$ for $\alpha=\frac{2}{5}$, and demonstrate that no two
solutions coincide. We will soon show that for any odd $q$ and even
$p$ the critical point $\omega_c$ does not exist.

Mathematically at critical points one of the coefficients of partial
fractions expansion, $A_k$ in Eq.~(\ref{Acoeff}), diverges. This
happens because when two $a_k$ coincide, $\hat{P}(s)$ can be written
as $\left(s-a_k\right)^2\hat{G}(s)$, where $\hat{G}(s)$ is some
polynomial in $s$ and $G(a_k)\neq 0$, and
\begin{equation}
 A_k=\frac{1}{\left(\displaystyle{ 2(s-a_k)\hat{G}(s)+ (s-a_k)^2\frac{d\hat{G}(s)}{ds}}\right)\displaystyle{\mid_{s=a_k}}}
\label{wc_3}
\end{equation}
diverges. And so in order to find such critical point two conditions must be satisfied
\begin{equation}
\hat{P}(s)=0,\qquad\frac{d\hat{P}(s)}{ds}=0. \label{wc_4}
\end{equation}
Using Eq.~(\ref{threeQuoter4}) with $\alpha=\frac{p}{q}$ and $\omega=\omega_c$,
\begin{equation}
\displaystyle{\left(s^2+{\omega_c}^2\right)^q+(-1)^{q-1}\gamma^qs^p=0}
\label{wc_5_01}
\end{equation}
\begin{equation}
\displaystyle{2qs\left(s^2+{\omega_c}^2\right)^{q-1}+p(-1)^{q-1}\gamma^qs^{p-1}=0}.
\label{wc_5_02}
\end{equation}
Solving these Eqs. we find
\begin{equation}
\displaystyle{s=\pm\sqrt{\frac{\alpha}{2-\alpha}}\omega_c}
\label{wc_6}
\end{equation}
and $\omega_c$ must satisfy
\begin{equation}
\displaystyle{\omega_c=\frac{1}{2^{\frac{1}{2-\alpha}}}}
\sqrt{(2-\alpha)\alpha^{\frac{\alpha}{2-\alpha}}}
\gamma^{\frac{1}{2-\alpha}}. \label{wc_7}
\end{equation}
Eq.~(\ref{wc_6}) and Eq.~(\ref{wc_7}) are only valid for even $q$ or
even $(q+p)$ (recall $\alpha=\frac{p}{q}$ and $\frac{p}{q}$ is
irreducible), since for any other case
Eqs.~(\ref{wc_5_01},\ref{wc_5_02}) has no solutions. The $+$ sign in
Eq.~(\ref{wc_6}) is for the case of even $q$ and odd $p$ and the $-$
sign for odd $q$ and $p$. To see this, insert the solution
Eq.~(\ref{wc_6}) in Eq.~(\ref{wc_5_01}) and then we must have
$(-1)^{q-1}<0$ and hence $q$ even, since $\frac{p}{q}$ is
irreducible $p$ is odd. Similarly for the $-$ solution in
Eq.~(\ref{wc_6}). From this discussion it becomes clear why there
are no critical frequency $\omega_c$ for $\alpha=\frac{2}{5}$
(Fig.~\ref{fig6}).

\begin{figure}
\begin{center}
\includegraphics[width=\columnwidth]{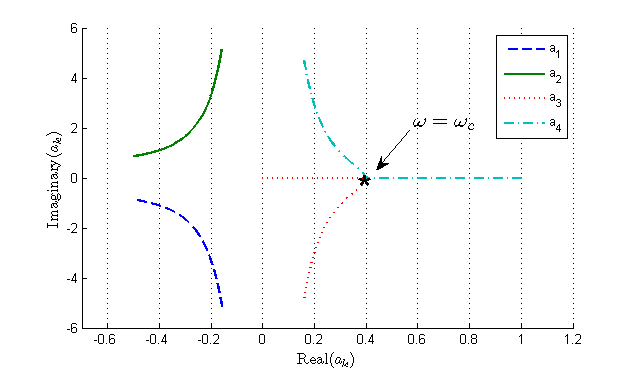}
\end{center}
\caption{ The four solutions of $\hat{P}(a_k)=0$ in the imaginary
plane for $\alpha=\frac{1}{2}$, $\gamma=1$ and $0<w<5$. At the
critical point $\omega=\omega_c=0.6873$ we have $a_3=a_4$. }
\label{fig5}
\end{figure}
\begin{figure}
\begin{center}
\includegraphics[width=\columnwidth]{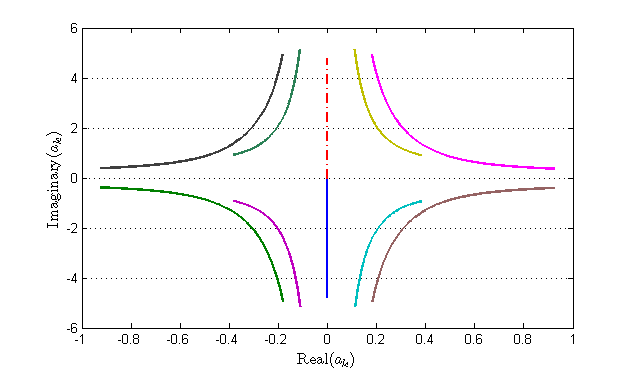}
\end{center}
\caption{The ten solutions of $\hat{P}(a_k)=0$ for
$\alpha=\frac{2}{5}$, and $\gamma=1$, in the imaginary plane, for
$0<w<5$. All ten solutions are different and do not coincide, namely
for this case there does not exist a critical frequency $\omega_c$
since $q=5$ is odd. The origin corresponds to $\omega=0$. }
\label{fig6}
\end{figure}

\subsection{Critical points $\omega_z$ and $\omega_m$}
\begin{figure}
\begin{center}
\includegraphics[width=\columnwidth]{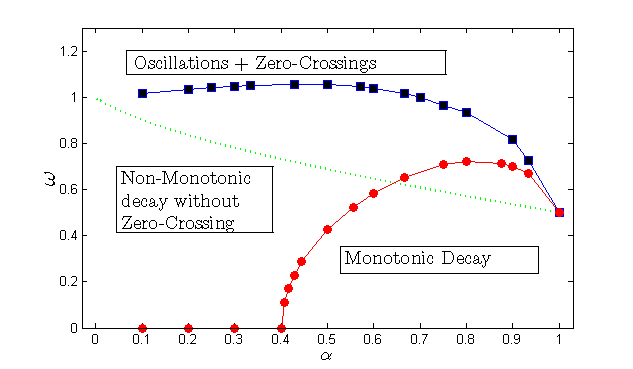}
\end{center}
\caption{  The phase diagram of the fractional oscillator.
Phase (a) monotonic decay of the correlation function
$C_x(t)$, phase (b) non-monotonic decay without zero-crossing and
(c) oscillations with zero crossings.
 The boundary between (b) and (c) is $\omega_z=\kappa_z(\alpha)$
 (solid line + squares), the boundary between (a) and (b) is
 $\omega_m=\kappa_m(\alpha)$ (solid line + circles).  For
$\alpha<\alpha_c\simeq 0.402$,
the phase of monotonic decay disappears, namely we do not find
over damped behavior.
 The dotted curve is the
 critical line $\omega_c$ given by Eq. (\ref{wc_7}).
All the curves are calculated for $\gamma=1$.
For $\alpha=1$, $\omega_c=\omega_z=\omega_m=\gamma/2$.
}
\label{fig7}
\end{figure}
We divide the phase space to three different regions. ($I$) $0<\omega<\omega_m$ the region where $C_x(t)$ decays monotonically ($II$) $\omega_m<\omega<\omega_z$ the region of non-monotonic decay while $C_x(t)$ always stays positive ($III$)
$\omega_z<\omega$ the region of non-monotonic decay while part of the time $C_x(t)$ is negative. 
 Similar to Eq.~(\ref{wc_7}) we find from dimensional analysis  
\begin{equation}
\omega_z=\kappa_z(\alpha)\gamma^{\frac{1}{2-\alpha}} \label{scaling2}
\end{equation}
and
\begin{equation}
\omega_m=\kappa_m(\alpha)\gamma^{\frac{1}{2-\alpha}} \label{scaling3}
\end{equation}
where $\kappa_z(\alpha)$ and $\kappa_m(\alpha)$ depend only on $\alpha$. By investigating analytical solution Eq.~(\ref{finalsol2}) for various $\alpha$ and $\gamma=1$ we obtain functions $\kappa_z(\alpha)$ and $\kappa_m(\alpha)$. The resulting phase diagram is presented in Fig. \ref{fig7}. One can readily see that $\omega_z$, $\omega_m$ and $\omega_c$ defined by Eq.~(\ref{wc_7}) all coincide only for the normal case $\alpha=1$.

An interesting behavior is observed for $\kappa_m(\alpha)$, as can be seen in Fig.~\ref{fig7} a sort of phase transition occurs around $\alpha\approxeq 0.4$. We used the general method developed in Sec.~\ref{GenSo} and explored the behavior of $\frac{1}{w^2}\frac{dC_x(t)}{dt}$, which in Laplace space is given by
\begin{equation}
\frac{1}{\omega^2}\left[s\hat{C}_x(s)-1\right]= -\frac{1}{s^2+\gamma
s^{\alpha}+\omega^2} . \label{w_c_add_01}
\end{equation}
For $\alpha\leq 0.4$ we always observed zero crossings even if we
decreased $\omega$ to $10^{-7}$ ($\gamma=1$), while for $\alpha\geq
0.404$ there were no zero crossings beneath some finite $\omega>0$, as is
shown in Fig.~\ref{fig7}. So we can conclude that there exist a
critical $\alpha$, $\alpha_c\approx 0.402\pm 0.002$, where for
$\alpha<\alpha_c$ $C_x(t)$ does not decay monotonically even if the
frequency of the binding Harmonic field $\omega\rightarrow 0$. Note that
the phase diagram Fig.~\ref{fig7} also exhibits some expected
behaviors: as we increase $\omega$ we find a critical line above which
the solutions are non-monotonic and exhibit zero crossing, a line
represented by $\kappa_z(\alpha)$.

To find accurate values of $\alpha_c$ we also used a method based on Bernstein theorem~\cite{Feller}. According to the theorem if and only if $f(t)$ is positive then for any integer $n$
\begin{equation}
0\leq(-1)^n\frac{d^n\tilde{f}(s)}{ds^n}
\qquad \left( s>0\right),
\label{new_met_01}
\end{equation}
where $\tilde{f}(s)$ is the Laplace pair of $f(t)$. As in the
previous paragraph in order to check the monotonicity of $C_x(t)$, we will inspect $\frac{1}{\omega^2}\frac{dC_x(t)}{dt}$
for zero crossings, or by speaking in the language of Bernstein
theorem by exploring $(-1)^n\frac{d^n\tilde{f}(s)}{ds^n}$, when
$-\tilde{f}(s)$ is given by Eq.~(\ref{w_c_add_01}). Using the
scaling relation of Eq.~(\ref{scaling3}) we can set $\gamma=1$, and
so it is easily checked that for $n=0$ and $n=1$
$(-1)^n\frac{d^n\tilde{f}(s)}{ds^n}>0$, for any $0<\alpha<1$. But
for $n=2$
\begin{equation}
\begin{array}{l}
\displaystyle{(-1)^n\frac{d^n\tilde{f}(s)}{ds^n}=
\frac{1}{(s^2+s^{\alpha}+\omega^2)^3}}
\\
\displaystyle{\qquad
\left( 6s^2+(9\alpha-\alpha^2-2)s^{\alpha}+(\alpha^2+\alpha)s^{2\alpha-2}\right.
}
\\
\displaystyle{ \qquad
\left.
-\omega^2(2+\alpha(\alpha-1))s^{\alpha-2}
\right),
}
\end{array}
\label{new_met_02}
\end{equation}
and in the limit $\omega\rightarrow 0$ one can easily show that for $\alpha\leqslant 0.071$ Eq.(\ref{new_met_02}) has negative values. Actually for $\alpha=0.01$ it is easily verified (by plotting Eq.(\ref{new_met_02})) that for any $\omega>0$ Eq.~(\ref{new_met_02}) would have negative values.
So for $n=2$ we have a upper bound for $\alpha$, $\alpha^{(2)}\approx 0.071$, where for any $\alpha<\alpha^{(2)}$ $\frac{dC_x(t)}{dt}$ crosses the zero line and the relaxation is non-monotonic. If one wishes to increase the accuracy for such an upper bound one should inspect the behavior of $(-1)^n\frac{d^n\tilde{f}(s)}{ds^n}$ for higher values of $n$, for any $n$ we define such a bound as $\alpha^{(n)}$.
Using Mathematica we can proceed to higher values of $n$ and find the upper bound $\alpha^{(n)}$ for different $\omega$. In Fig.~\ref{fig8} we plotted the upper bounds as a function of $n$ for various $\omega$, we see that as $n$ grows the upper bound $\alpha^{(n)}$ converges to some value $<1$. The results achieved by this method converge to values very close to the values displayed in Fig.~\ref{fig7}, for example for $\omega=0$ and $n=150$ $\alpha^{(150)}=0.394$, compared with $\alpha_c\approxeq 0.402$, obtained by inspection of exact solution.

A physical explanation for this interesting result is based on the cage effect. For small $\alpha$ the friction force induced by the medium is not just slowing down the particle but also causing the particle a rattling motion. To see this consider the FLE Eq.~(\ref{fracGen}) in the limit $\alpha\rightarrow 0$
\begin{equation}
m\ddot{x}+m\gamma(x-x_0)+m\omega^2x\approx \xi(t) \label{phys_expl01}
\end{equation}
where $x_0$ is the initial condition. Eq.(\ref{phys_expl01})
describes harmonic motion and the friction $\gamma$ in this
$\alpha\rightarrow 0$ limit yields an elastic harmonic force. In
this sense the medium is binding the particle preventing diffusion
but forcing oscillations. In the opposite limit of
$\alpha\rightarrow 1$
\begin{equation}
m\ddot{x}+m\gamma\dot{x}+m\omega^2x\approx \xi(t), \label{phys_exp02}
\end{equation}
the usual damped oscillator with noise is found. So from Eq.~(\ref{phys_expl01}) an oscillating behavior is expected, even when $\omega\rightarrow 0$ which can be explained by the rattling motion of a particle in the cage formed by the surrounding particles. This behavior manifests in the non-monotonic oscillating solution we have found for small $\alpha$. Our findings that $\alpha_c$ marks a non smooth transition between normal friction $\alpha\rightarrow 1$ and elastic friction $\alpha\rightarrow 0$ is certainly a surprising result.

\begin{figure}
\begin{center}
\includegraphics[width=\columnwidth]{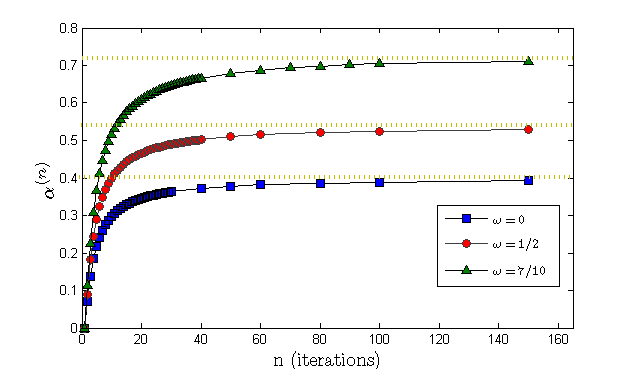}
\end{center}
\caption{ Using the Bernstein Method we find  the upper bound $\alpha^{(n)}$ as a function of $n$, the number of the derivatives. We consider three different $\omega$ ($\omega=0$, $\omega=\frac{1}{2}$ and $w=\frac{7}{10}$). The dashed lines present the values found numerically in Fig.~\ref{fig7}. As $n$ grows a convergence of the bound is achieved as is seen for $\omega=0$ as $\alpha^{(n)}$ converges to $\alpha_c\approxeq 0.402\pm 0.002$.
}
\label{fig8}
\end{figure}

\section{Response to an external field}
\label{response_00}
In this section we will explore the response of the FLE to an external time dependent force. The response of a system to an oscillating time dependent field naturally leads to the phenomena of resonances, when the frequency of the external field matches a natural frequency of the system. The response of sub-diffusing systems to such time dependent field was the subject of intensive research~\cite{Barbi,Sokolov2,Goychuk1}. In particular fractional approach to sub-diffusion naturally leads to anomalous response functions commonly found in many systems e.g. the Cole-Cole relaxation~\cite{Barkai3,Weron,Goychuk2}. In the next three subsections we will explore the response of FLE with and without the Harmonic potential and also investigate the behavior of the imaginary part of the complex susceptibility, i.e. the dielectric loss, for these cases.

Our starting point is Eq.~(\ref{int1}) with $F(x,t)=F_0\cos(\Omega t)-m\omega^2x$, performing an average we obtain
\begin{equation}
\ddot{\langle x\rangle}+\gamma\frac{d^{\alpha}\langle x\rangle}{dt^{\alpha}}
+\omega^2\langle x \rangle=
\frac{F_0}{m}\cos(\Omega t).
\label{Sec1_02}
\end{equation}
The solution in the long time regime is
\begin{equation}
\langle x(t)\rangle \sim
\frac{F_0}{m}\int_{0}^{t}\cos(\Omega (t-\grave{t}))h(\grave{t})\,d\grave{t},
\label{Sec1_03}
\end{equation}
where $h(t)$ is soon defined. Eq.~(\ref{Sec1_03}) could be written as
\begin{equation}
\langle x(t)\rangle=
R(\Omega)\cos\left(\Omega t+\theta(\Omega)\right)\qquad t \rightarrow \infty.
\label{Sec1_06}
\end{equation}
The response $R(\Omega)$ and the phase shift $\theta(\Omega)$ are obtained by the means of complex susceptibility
\begin{equation}
\chi(\Omega)=\chi'(\Omega)+i\chi''(\Omega)=
\hat{h}(-i\Omega)
\label{Sec1_09}
\end{equation}
where $\chi'(\Omega)$ and $\chi''(\Omega)$ are the complex and the imaginary parts of the susceptibility, respectively, and $\hat{h}(-i\Omega)=\int_o^\infty e^{i\Omega t}h(t)\,dt$~\cite{Kubo1}. For the response
\begin{equation}
R(\Omega)=|\chi(\Omega)|
\label{Sec1_07}
\end{equation}
and
\begin{equation}
\theta(\Omega)=
\arctan\left(-\frac{\chi''(\Omega)}{\chi'(\Omega)}\right),
\label{Sec1_08}
\end{equation}
for the phase shift.

\subsection{Unbounded Particle}
\label{Free_Motion}

For the unbounded particle we set $\omega=0$ in Eq.~(\ref{Sec1_02}) and using Eq.~(\ref{recLap1}) we obtain for $\chi(\Omega)$
\begin{equation}
\chi(\Omega)=\hat{h}(-i\Omega)=\frac{1}{\gamma(-i\Omega)^{\alpha}-\Omega^{2}}.
\label{unbound01}
\end{equation}
By the use of Eq.~(\ref{Sec1_07}) we can explicitly obtain now the behavior of $R(\Omega)$ for any $0<\alpha<1$.
For the normal diffusion case when $\alpha=1$, the response $R(\Omega)$ is a decaying function of $\Omega$ and no resonance is observed, but the picture is quite different for $0<\alpha<1$. In this sub-diffusive part the response $R(\Omega)$ is not always a monotonically decaying function and could obtain a maximum, i.e. a resonance, even for such a free motion. As is seen in Fig.~\ref{fig9} for small enough $\alpha$, $R(\Omega)$ has a resonance, we will show that the existence of the resonance does not depend on any other parameter but $\alpha$.
\begin{figure}
\begin{center}
\includegraphics[width=\columnwidth]{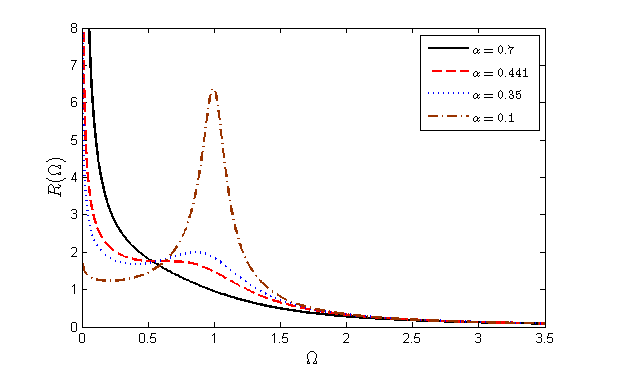}
\end{center}
\caption{
The response of FLE to an oscillating field for a free particle and different $\alpha$'s. For $\alpha<\alpha_R$ a resonance is observed. All the curves are plotted with $\gamma=1$.
}
\label{fig9}
\end{figure}

Writing down the response $R(\Omega)$ explicitly
\begin{equation}
R(\Omega)=
\displaystyle{\frac{1}{\sqrt{\Omega^4+\gamma^2 \Omega^{2\alpha}-2\gamma \Omega^{2+\alpha}\cos\left(\frac{\pi\alpha}{2}\right)}}},
\label{Sec1_10}
\end{equation}
We are looking for the solutions of $dR(\Omega)/d\Omega=0$, and hence
\begin{equation}
4\Omega_R^3+2\alpha \gamma^2 \Omega_R^{2\alpha-1}
-2(\alpha+2)\gamma\Omega_R^{\alpha+1}\cos\left(\frac{\pi\alpha}{2}\right)=0,
\label{Sec1_12}
\end{equation}
so the solution is
\begin{equation}
\begin{array}{l}
\displaystyle{\frac{\gamma}{\Omega_R^{2-\alpha}}} \\
=\frac{1}{2\alpha}
\left((\alpha+2)\cos\left(\frac{\pi\alpha}{2}\right)\pm
\sqrt{(\alpha+2)^2\cos^2\left(\frac{\pi\alpha}{2}\right)-8\alpha}\right),
\end{array}
\label{Sec1_13}
\end{equation}
and $\Omega_R$ is the frequency for which the resonance is found. When the discriminant on the right hand of Eq.~(\ref{Sec1_13}) is greater than zero
there will be a resonance . The discriminant has no dependence on $\gamma$, and is always positive for $\alpha\leq\alpha_R=0.441021...$ which satisfies the following relation
\begin{equation}
(\alpha_R+2)^2\cos^2\left(\frac{\pi\alpha_R}{2}\right)-8\alpha_R=0
\label{Sec1_15}
\end{equation}
So no resonance is found for $\alpha>\alpha_R$ and  for $\alpha<\alpha_R$ there is always exist an $\Omega_R>0$ for which the response will exhibit a resonance. This result of a resonance for a free particle FLE is highly unexpected but sets well with our description of the friction force for small $\alpha$'s as an elastic force due to the cage effect.
\begin{figure}
\begin{center}
\includegraphics[width=\columnwidth]{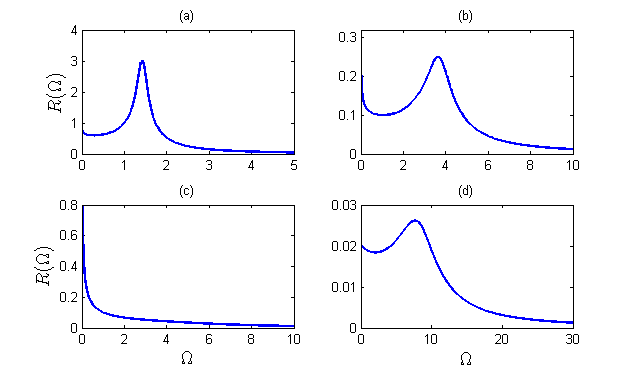}
\end{center}
\caption{
The response of FLE to an oscillating field for a harmonically bounded particle and different $\alpha$s. (a) $\alpha=0.2$, $\omega=1$ and $\gamma=1$. (b) $\alpha=0.2$, $\omega=1$ and $\gamma=10$. (c) $\alpha=0.7$, $\omega=1$ and $\gamma=10$. (d) $\alpha=0.7$, $\omega=7$ and $\gamma=10$.
}
\label{fig10}
\end{figure}

\subsection{Harmonically Bounded Particle}
\label{Harmonic_b01}
 Now we treat the response function of the FLE with a harmonic field, i.e. of a fractional oscillator.
Starting with Eq.~(\ref{Sec1_02}) we set the initial conditions $x_0=v_0=0$ and in the long time limit $t\rightarrow\infty$ we  obtain again $\langle x \rangle=R(\Omega)\cos(\Omega t+\theta(\Omega))$ where $R(\Omega)$ and $\theta(\Omega)$ are defined by Eqs.~(\ref{Sec1_07},\ref{Sec1_08}). The complex susceptibility for such case is obtained
\begin{equation}
\chi(\Omega)=\frac{1}{(\omega^2-\Omega^2)
+\gamma\left(-i\Omega\right)^{\alpha}}.
\label{Sec2_06}
\end{equation}
Eq.~(\ref{Sec2_06}) was already obtained~\cite{Coffey1,Coffey2,Coffey3} for the fractional Klein-Kramers equation~\cite{Barkai2} in the high damping limit and is called a generalized Rocard equation~\cite{Rocard,Scaife}. For $\alpha=1$ we find the complex susceptibility of a normal damped oscillator.

We are interested in the resonance points for the response to the applied field, i.e. points of maximum of $R(\Omega)$, which generally depend on $\Omega$, $\gamma$ and $\omega$ (see Fig.~\ref{fig10}). For the normal oscillator there is a resonance if the condition $\omega\geq\frac{1}{\sqrt{2}}\gamma$ is satisfied. If this condition is not satisfied, $R(\Omega)$  is a monotonically decreasing function of $\Omega$. From Eqs.~(\ref{Sec1_07},\ref{Sec2_06})
\begin{equation}
R(\Omega)=\frac{1}{\sqrt{(\omega^2-\Omega^2)^2+\gamma^2 \Omega^{2\alpha}
+2(\omega^2-\Omega^2) \gamma \Omega^\alpha \cos\left(\frac{\pi\alpha}{2}\right)}},
\label{Sec2_07}
\end{equation}
and using $dR(\Omega)/d\Omega=0$, we find
\begin{equation}
\begin{array}{l}
2\alpha(\frac{\gamma}{\Omega_R^{2-\alpha}})^2
+[2\alpha \left(\frac{\omega^2}{\Omega_R^2}-1\right) -4]\cos\left(\frac{\pi\alpha}{2}\right) \frac{\gamma}{\Omega_R^{2-\alpha}}\\
\qquad\qquad\qquad\qquad\qquad\qquad
 -4\left(\frac{\omega^2}{\Omega_R^2}-1\right)=0.
\end{array}
\label{Sec2_09}
\end{equation}
\begin{figure}
\begin{center}
\includegraphics[width=\columnwidth]{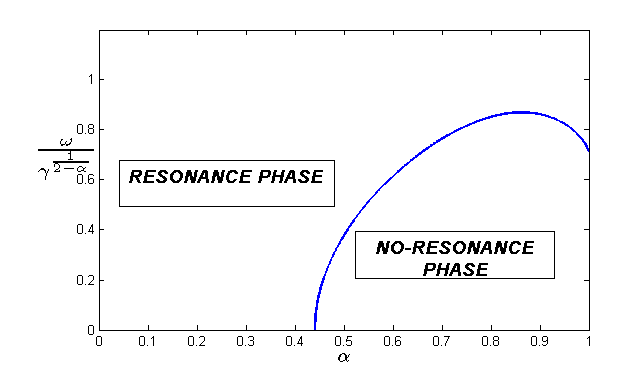}
\end{center}
\caption{
Phase diagram of the response of the system with a harmonic field
to a oscillating time dependent force field.
Two simple behaviors are found either a resonance exists (``Resonance Phase")
or not (``No-Resonance Phase").
For $\alpha<\alpha_{R}=0.441...$ a resonance exists for any
binding harmonic field and any friction $\gamma$.
}
\label{fig11}
\end{figure}

 The exploration of Eq.~(\ref{Sec2_09}) (see Appendix~\ref{AppB}) leads to two findings which are presented in Fig.~\ref{fig11}. The first one is the existence of the same critical $\alpha_R$ given by Eq.~(\ref{Sec1_15}) for the response of FLE with a harmonic potential, i.e. for any $\alpha<\alpha_R$ there always exist a specific $\Omega_R$ (which depends on $\gamma$ and $\omega$) for which the system is in resonance. The second finding is that above $\alpha_R$ we have a well defined boundary between the phase for which the resonance exist - a ``Resonance Phase", and a phase where are no resonance - ``No-Resonance Phase" (see Fig.~\ref{fig11}). For $\alpha<\alpha_R$ ``No-Resonance-Phase" does not exists. The boundary is given by the following relation
\begin{equation}
\displaystyle{\frac{\omega}{\gamma^{\frac{1}{2-\alpha}}}=g(\alpha)},
\label{Sec2_16}
\end{equation}
and the phase diagram is presented in Fig.~\ref{fig11}. The function $g(\alpha)$ is found analytically (see Appendix~\ref{AppB})
\begin{equation}
g(\alpha)=
{c(\alpha)}^{-\frac{1}{2-\alpha}}
\sqrt{\frac{{c(\alpha)}^2-c(\alpha)\cos\left(\frac{\pi\alpha}{2}\right)(1+\frac{2}{\alpha})
+\frac{2}{\alpha}}
{\frac{2}{\alpha}-c(\alpha)\cos\left(\frac{\pi\alpha}{2}\right)}}
\label{Sec2_23}
\end{equation}
where $c(\alpha)$ is given by Eq.~(\ref{qwerf}). As is seen from Fig.~\ref{fig11} for ${\omega}/{\gamma^{\frac{1}{2-\alpha}}}>g(\alpha)$ a ``Resonance Phase'' is obtained and ``No-Resonance Phase'' for ${\omega}/{\gamma^{\frac{1}{2-\alpha}}}<g(\alpha)$. In the limit of $\alpha\rightarrow 1$ the boundary goes to the expected value for damped oscillator $1/\sqrt{2}$. Near the critical point $\alpha_R$ the $g(\alpha)$ drops to zero as a power-low with exponent $1/2$
\begin{equation}
g(\alpha)\propto(\alpha-\alpha_R)^{\frac{1}{2}}\qquad \alpha\rightarrow \alpha_R.
\label{res_exp}
\end{equation}

The existence of the same critical $\alpha_R$ for a free and a harmonically bounded particle is easily understood from the phase diagram in Fig.~\ref{fig11}. Choosing the straight line $\omega=0$, which represent the free particle, and going along this line when starting at the ``Resonance Phase" we will cross to the ``No-Resonance Phase" exactly for $\alpha=\alpha_R$. Generally speaking the same critical $\alpha_R$ will be obtained for any kind of external force because it is determined by the internal properties of the surrounding medium represented by the friction part in FLE.  The phase diagram in Fig.~\ref{fig11} has much in common with the phase diagram in Fig.~\ref{fig7}, which is quite reasonable because of the strong connection between non-monotonic decay of the correlation $C_x(t)$ and existence of a resonance. A presence of a resonance for small enough $\alpha$ emphasize the previously obtained result of non-existence of the over-damped limit for such $\alpha$. Those properties are due to the same cage effect that we already discussed.

\subsection{Complex Susceptibility}
\label{complex_01}

Eq.~(\ref{Sec2_06}) for the complex susceptibility can be written in the following form
\begin{widetext}
\begin{equation}
\begin{array}{c}
\displaystyle{
\chi(\Omega)=\chi '(\Omega)+i\chi ''(\Omega)=
}\\
\displaystyle{
\frac{(\omega^2-\Omega^2)+\gamma\Omega^{\alpha}\cos\left(\frac{\pi\alpha}{2}\right)}
{(\omega^2-\Omega^2)^2+\gamma^2\Omega^{2\alpha}+2\gamma(\omega^2-\Omega^2)\Omega^{\alpha}
\cos\left(\frac{\pi\alpha}{2}\right)}
+i\frac{\gamma\Omega^{\alpha}\sin\left(\frac{\pi\alpha}{2}\right)}
{(\omega^2-\Omega^2)^2+\gamma^2\Omega^{2\alpha}+2\gamma(\omega^2-\Omega^2)\Omega^{\alpha}
\cos\left(\frac{\pi\alpha}{2}\right)}
},
\end{array}
\label{Sec3_01}
\end{equation}
\end{widetext}
the real and the imaginary parts of the susceptibility are experimentally measured quantities for many systems and so it is interesting to explore their behavior for the FLE. In this subsection we will explore the behavior of the imaginary part $\chi ''(\Omega)$ which is also called ``the loss". From Fig.~\ref{fig13} we observe interesting behavior of $\chi ''(\Omega)$ for different $\alpha$'s, not only one peak is present as is expected for the normal oscillator, but we observe a double peak phenomena for some $\alpha$'s. The double peak phenomena of ``the loss" for super-cooled liquids and protein solutions is a well known phenomena~\cite{Dachwitz,Gotze,Bagchi} and usually treated by the means of mode-coupling theory~\cite{Gotze}. We define two phases for the behavior of $\chi ''(\Omega)$  and in the following find the phase diagram for $\chi ''(\Omega)$. The first phase is the phase where $\chi ''(\Omega)$ has only one peak - ``One-Peak" phase and a ``Double-Peak" phase, where a double peak phenomena is observed.
\begin{figure}
\begin{center}
\includegraphics[width=\columnwidth]{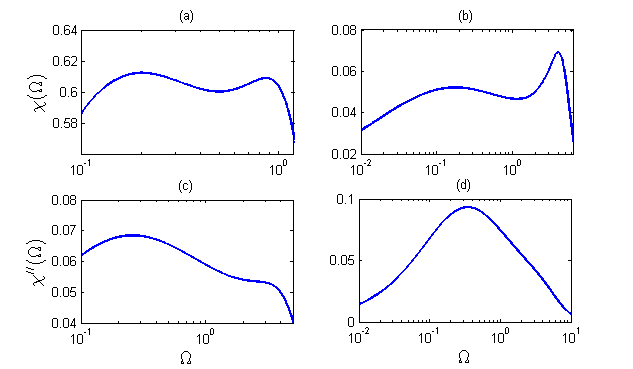}
\end{center}
\caption{
The imaginary part of the complex susceptibility for various $\alpha$. A double peak phenomena is observed. (a) $\alpha=0.66$, $\gamma=1.8$ and $\omega=0.7$. (b) $\alpha=0.5$, $\gamma=10$ and $\omega=2$. (c) $\alpha=0.63$, $\gamma=10$ and $\omega=2$. (d) $\alpha=0.8$, $\gamma=10$ and $\omega=2$.
}
\label{fig13}
\end{figure}
\begin{figure}
\begin{center}
\includegraphics[width=\columnwidth]{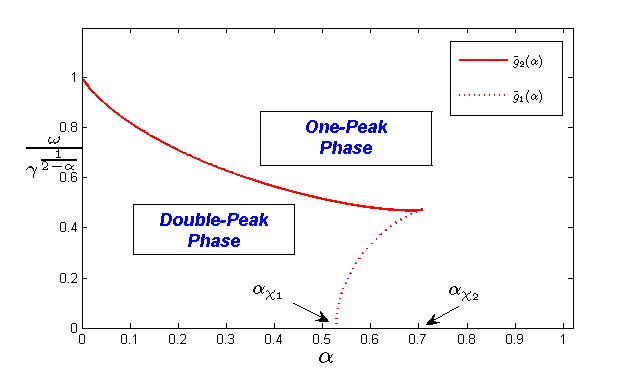}
\end{center}
\caption{
Phase diagram of the imaginary part of complex susceptibility.
Two phases are found, a phase with a presence of one maxima for $\chi ''(\Omega)$ - ``One-Peak" phase, or a phase with a presence of two maxima - ``Double-Peak" phase. For $\alpha>\alpha_{\chi_2}=0.707...$ only one phase present.
}
\label{fig14}
\end{figure}

In Appendix~\ref{AppC} we explored $d\chi ''(\Omega)/d\Omega$ and
searched for the boundaries between the ``One-Peak" and
``Double-Peak" phases. The result is presented in Fig.~\ref{fig14}.
The boundaries between the phases are given by analytical functions
$\tilde{g}_1(\alpha)$ and $\tilde{g}_2(\alpha)$ which are only
$\alpha$ dependent. Two critical $\alpha$ are found for such a phase
diagram, the first one $\alpha_{\chi_1}=0.527...$ for which the
boundary $\tilde{g}_1(\alpha)$ drops to zero, and a second one
$\alpha_{\chi_2}=0.707...$ for which $\tilde{g}_1(\alpha)$ and
$\tilde{g}_2(\alpha)$ coincide (see Fig.~\ref{fig13}). We
also must note that near $\alpha_{\chi_1}$ $\tilde{g}_1(\alpha)$
behaves like $\tilde{g}_1(\alpha)\propto (\alpha-\alpha_{\chi
1})^{1/2}$, a behavior which was also observed for $g(\alpha)$ and
$\kappa_m(\alpha)$ near the corresponding critical points.

The double-peak phenomena in our case is explained in the same sense as was explained the existence of resonance for small enough $\alpha$, and the disappearance of the monotonic decay phase for the correlation function. The reason is the same, the friction becomes more of an elastic force for such small $\alpha$ embedding oscillations in the system. Such a claim is emphasized by the Cole-Cole plots of the complex susceptibility, presented in Fig.~\ref{fig15}. For $\alpha=0.8$ the behavior is very much as for a normal damped oscillator, a Debye model~\cite{Kubo1}, for small enough $\omega/\gamma^{1/(2-\alpha)}$ (large friction) - Fig.~\ref{fig15}.(c), i.e. a monotonic behavior for the relaxation, and a Van Vleck-Weisskopf-Fr\"{o}hlich type~\cite{Kubo1} for a large value of $\omega/\gamma^{1/(2-\alpha)}$ (small friction)- Fig.~\ref{fig15}.(d), i.e. a oscillating behavior of the relaxation.
These two prototypes of normal complex susceptibility correspond to presence of a single characteristic frequency in the system, or a single time scale if we are concerned with correlations. For small $\alpha$, like $\alpha=0.1$ in Fig.~\ref{fig15}(a), we see a coexistence of these two types of normal susceptibilities. The right side of Fig.~\ref{fig15}(a) corresponds to a Debye type - a monotonically decaying process and on the left side a Van Vleck-Weisskopf-Fr\"{o}hlich type wich shows highly oscillating behavior even in the case when $\gamma$ and $\omega$ are the same as for Fig.~\ref{fig15}.(c). Effectively for small $\alpha$ we have two characteristic frequencies in the system, the lowest is responsible for the monotonic decay and the high frequency is an oscillating process, for intermediate $\alpha$ we have some mixed behavior - Fig.~\ref{fig15}.(b). This oscillating behavior that is seen in Fig.~\ref{fig15}.(a) is the manifestation of the cage effect which we already explained as the rattling motion of the surrounding particles and is presented in FLE due to the friction force.

\begin{figure}
\begin{center}
\includegraphics[width=\columnwidth]{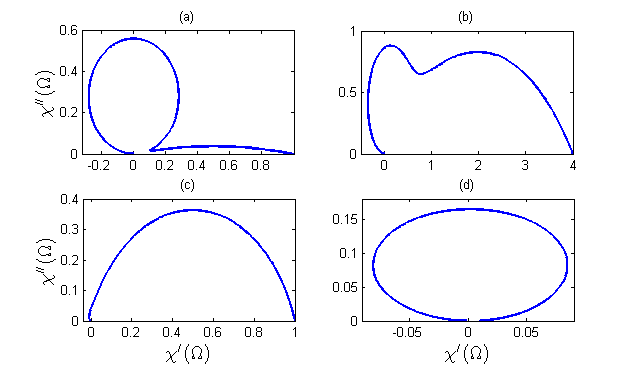}
\end{center}
\caption{
The Cole-Cole plots of the complex susceptibility, $\chi ''(\Omega)$ as a function of $\chi '(\Omega)$. (a) $\alpha=0.1$, $\gamma=10$ and $\omega=1$. (b) $\alpha=0.5$, $\gamma=10$ and $\omega=0.5$. (c) $\alpha=0.8$, $\gamma=10$ and $\omega=1$. (d) $\alpha=0.8$, $\gamma=1$ and $\omega=10$. Panel (c) and (d) show two typical normal behaviors,  for small $\alpha$ the panels (a) and (b) show a behavior which is a mixture of the two normal typical behaviors.
}
\label{fig15}
\end{figure}

\section{Summary}
\label{Summary}
The Fractional Langevin Equation (FLE) with power-law memory kernel and $0<\alpha<1$ is a stochastic framework
describing anomalous sub-diffusive behavior. This equation may be
expressed in terms of fractional derivative and so provides an example of a physical phenomena where
non-integer calculus plays a central role. The solution of a fractional-differential equation describing the correlation function was
presented in terms of roots of regular polynomials. 
It was shown that for $\alpha\neq 1$ there is no unique way to
define an over-damped or under-damped motion. Three definitions were
proposed for the frequency of transition, i.e. $\omega_c$,
$\omega_m$ and $\omega_z$. We observed an existence of a phase transition for a
critical $\alpha=\alpha_c\approx 0.402$, where for $\alpha<\alpha_c$
$C_x(t)$ does not decay monotonically for any $\omega>0$. Physically
it is explained using a cage effect: a rattling motion of a particle
in the cage formed by the surrounding particles. A response to a
time dependent field in terms of complex susceptibility $\chi
(\Omega)$ also was calculated and similar critical $\alpha$'s were
found. Particularly for $\alpha<\alpha_R=0.441...$ the system will
always be in a resonance with the external field for particular
$\Omega$ and any $\gamma$ and $\omega$, even in the case of free
particle ($\omega=0$). For the ``loss" - the imaginary part of the
complex susceptibility, $\chi ''(\Omega)$, two phases were defined:
(i)``One-Peak" phase were the complex susceptibility obtains only
one maximum as in a regular case, (ii) ``Double-Peak" phase, were
the complex susceptibility obtains two maxima, a phase diagram was
presented. Two critical exponents $\alpha_{\chi_1}=0.527...$ and
$\alpha_{\chi_2}=0.707...$ were found for $\chi ''(\Omega)$,
exponents which define the boundaries of the phase diagram. In
conclusion, critical exponents like $\alpha_c$, $\alpha_R$,
$\alpha_{\chi_1}$ and $\alpha_{\chi_2}$, mark sharp transitions in
the behaviors of systems with fractional dynamics. Thus, these
critical exponents are clearly important and general in the
description of anomalous kinetics.

{\bf Acknowledgment} This work was supported by the Israel Science
Foundation.


\section{Appendix A: The solution for non-distinct zeros of $\hat{P}(s)$}
{\label{appa_a}} In this Appendix we derive the solution of
Eq.~(\ref{comp2}) for the case when two zeros of $\hat{P}(s)$
coincide. This means that $\hat{P}(s)$ has $2q-2$ distinct zeros of
order $1$ and one zero of order $2$. 
Namely, at the critical point $\omega_c$ only two $a_k$ coincide and we
present here a method of solution for $\omega=\omega_c$.

Starting with Eq.~(\ref{comp2}), we write the partial fraction expansion in the following way
\begin{equation}
\frac{1}{\hat{P}(s)}=\sum_{k=1}^{2q-1}\frac{A_k}{s-a_k}\,+\,
\frac{\tilde{A}}{(s-a_{2q-1})^2} \label{app01}
\end{equation}
where $a_k$ are the zeros of $\hat{P}(s)$ and we assign $a_{2q-1}$
to be the zero of the second order. $A_k$ and $\tilde{A}$ are given
by
\begin{equation}
A_k=\frac{1}{\displaystyle\frac{d\hat{P}(s)}{ds}\mid_{s=a_k}}\qquad
1\leq k<2q-1, \label{app02}
\end{equation}
\begin{equation}
A_{2q-1}=-\sum_{k=1}^{2q-2}A_k,
\label{app03}
\end{equation}
and
\begin{equation}
\tilde{A}=\frac{1}{\displaystyle\frac{d}{ds}\frac{\hat{P}(s)}{s-a_{2q-1}}
\mid_{s=a_{2q-1}}}. \label{app04}
\end{equation}
Using the relation~\cite{Miller}
\begin{equation}
\sum_{k=1}^{2q-1}a_k^mA_k\,+\,ma_k^{m-1}\tilde{A}=0\qquad m=0,1,\dots ,2q-2
\label{app05}
\end{equation}
one finds that
\begin{equation}
\begin{array}{l}
\displaystyle{\frac{s^m}{\hat{P}(s)}=\sum_{k=1}^{2q-1}\frac{A_ka_k^m}{s-a_k}\,+\,
\frac{ma_{2q-1}^{m-1}\tilde{A}}{s-a_{2q-1}}\,+\,
\frac{a_{2q-1}^{m}\tilde{A}}{(s-a_{2q-1})^2}}\\
\qquad\qquad\qquad\qquad\qquad\qquad m=0,1,\dots,2q-1.
\end{array}
\label{app06}
\end{equation}
Hence using Eqs.~(\ref{comp2},\ref{qmult},\ref{app06})
\begin{widetext}
\begin{equation}
\hat{C}_x(s)=\sum_{m=0}^{2q-1}\sum_{j=0}^{q-1}\left(
\sum_{k=1}^{2q-1}\frac{a_k^mA_k\tilde{B}_{mj}}{s-a_k}s^{-\frac{j}{q}}\,+\,
\frac{ma_{2q-1}^{m-1}\tilde{A}\tilde{B}_{mj}}{s-a_{2q-1}}s^{-\frac{j}{q}}\,+\,
\frac{a_{2q-1}^{m}\tilde{A}\tilde{B}_{mj}}{(s-a_{2q-1})^2}s^{-\frac{j}{q}}
\right), \label{app07}
\end{equation}
and it is only left to perform an Inverse Laplace Transform of $\displaystyle\frac{1}{s^{\frac{j}{q}}{(s-a_{2q-1})^2}}$, using convolution theorem
\begin{equation}
\displaystyle\frac{1}{s^{\frac{j}{q}}{(s-a_{2q-1})^2}}\;\Laplace\quad
\frac{te^{a_{2q-1}t}}{\Gamma(\frac{j}{q})a_{2q-1}^{\frac{j}{q}}}
\gamma(\frac{j}{q},a_{2q-1}t)-
\frac{e^{a_{2q-1}t}}{\Gamma(\frac{j}{q})a_{2q-1}^{\frac{j}{q}+1}}
\gamma(\frac{j}{q}+1,a_{2q-1}t)
\label{app08}
\end{equation}
or using Mittag-Leffler function
\begin{equation}
\displaystyle\frac{1}{s^{\frac{j}{q}}{(s-a_{2q-1})^2}}\;\Laplace\quad
t^{\frac{j}{q}+1}E_{1,1+\frac{j}{q}}\left(a_{2q-1}t\right)-
\frac{j}{q}t^{\frac{j}{q}+1}E_{1,2+\frac{j}{q}}\left(a_{2q-1}t\right).
\label{app09}
\end{equation}
Finally using Eqs.~(\ref{type2sol},\ref{app07},\ref{app08})
\begin{equation}
C_x(t)=\sum_{m=0}^{2q-1}\sum_{j=0}^{q-1}\frac{\tilde{B}_{mj}}{\Gamma(\frac{j}{q})}
\left[ \sum_{k=1}^{2q-1}
a_k^{m-\frac{j}{q}}A_ke^{a_kt}
\gamma(\frac{j}{q},a_kt)+
a_{2q-1}^{m-\frac{j}{q}}\tilde{A}e^{a_{2q-1}t}
\left((t+ma_{2q-1}^{-1})\gamma(\frac{j}{q},a_{2q-1}t)
-a_{2q-1}^{-1}\gamma(\frac{j}{q}+1,a_{2q-1}t)\right)
\right],
\label{app10}
\end{equation}
or using Eq.~(\ref{app09})
\begin{equation}
C_x(t)=\sum_{m=0}^{2q-1}\sum_{j=0}^{q-1}\tilde{B}_{mj}t^{\frac{j}{q}}
\left[ \sum_{k=1}^{2q-1}
a_k^{m}A_k
E_{1,1+\frac{j}{q}}(a_kt)+
a_{2q-1}^{m}\tilde{A}
\left((t+ma_{2q-1}^{-1})E_{1,1+\frac{j}{q}}(a_{2q-1}t)
-\frac{j}{q}tE_{1,2+\frac{j}{q}}(a_{2q-1}t)\right)
\right].
\label{app11}
\end{equation}
\end{widetext}

A final remark: one can show that for our case of integer $q>p>0$, third and higher order zeros of $\hat{P}(s)$ don't exist.

\section{Appendix B: Exploration of Eq.~(\ref{Sec2_09})}
\label{AppB}
In this Appendix we prove the existence of $\alpha_R$ for FLE with a harmonic force and derive the equation for $g(\alpha)$ given by Eq.~(\ref{Sec2_23}).
 Solving Eq.~(\ref{Sec2_09}) one gets for $\displaystyle{\gamma/\Omega_R^{2-\alpha}}$
\begin{equation}
\begin{array}{l}
\displaystyle{\frac{\gamma}{\Omega_R^{2-\alpha}}=\frac{1}{2\alpha}\left([2-\alpha y]
\cos\left(\frac{\pi\alpha}{2}\right)\right)}\\
\displaystyle{\pm\frac{1}{2\alpha}\sqrt{\left([2-\alpha y]^2
\cos\left(\frac{\pi\alpha}{2}\right)\right)^2+8\alpha y}}=q_{1\pm}(y),
\end{array}
\label{Sec2_10}
\end{equation}
and $\displaystyle{y=\left(\frac{\omega^2}{\Omega_R^2}-1\right)}>-1$.
Writing the left hand side of Eq.~(\ref{Sec2_10}) in terms of $y$
\begin{equation}
\frac{\gamma}{\Omega_R^{2-\alpha}}=\frac{\gamma}{\omega^{2-\alpha}}(y+1)^{1-\frac{\alpha}{2}}
=q_2(y).
\label{Sec2_11}
\end{equation}
We see that for the extrema points of $R(\Omega)$ the functions $q_{1\pm}(y)$ and $q_2(y)$ cross each other (see Fig.~\ref{fig12}). While $q_2(y)$ is a monotonic increasing function starting from zero for $y=-1$ and growing as $y^{1-\frac{\alpha}{2}}$ for large $y$, $q_{1\pm}(y)$ constructs two branches  where $q_{1+}(y)$ is the upper branch and $q_{1-}(y)$ is the lower branch and for some point $y_\alpha$
\begin{equation}
q_{1-}(y_\alpha)=q_{1+}(y_\alpha)=
\frac{(2-\alpha y_\alpha)\cos\left(\frac{\pi\alpha}{2}\right)}{2\alpha}.
\label{Sec2_12}
\end{equation}
If $y_\alpha<-1$ then $q_2(y)$ crosses $q_{1\pm}(y)$ no mater what the parameters $\gamma$,$\omega$ and $\alpha$ are, because in that case for $y=-1$ $q_{1\pm}(y)>0$ and $q_2(y)=0$ and a resonance is always obtained. The point $y_\alpha$ is derived from Eq.~(\ref{Sec2_10}) and determined by the following relation
\begin{equation}
(2-\alpha y_\alpha)^2\cos^2\left(\frac{\pi\alpha}{2}\right)+8\alpha y_\alpha=0.
\label{Sec2_13}
\end{equation}
Solving Eq.~(\ref{Sec2_13}) in terms of $y_\alpha$ one finds
\begin{equation}
y_\alpha=-\frac{2}{\alpha\cos^2\left(\frac{\pi\alpha}{2}\right)}
\left(1-\sin\left(\frac{\pi\alpha}{2}\right)\right)^2,
\label{Sec2_14}
\end{equation}
where we took the $-$ sign because $y>-1$. For $0<\alpha<1$ Eq.~(\ref{Sec2_14}) is an increasing function of $\alpha$ and so we have a critical $\alpha$, $\alpha_R$ for $y_\alpha=-1$. Eq.~(\ref{Sec2_13}) with $y_\alpha=-1$ is exactly Eq.~(\ref{Sec1_15}) which defines the equation for $\alpha_R$ and so we have shown the existence of $\alpha_R$ for the harmonically bounded particle.

\begin{figure}
\begin{center}
\includegraphics[width=\columnwidth]{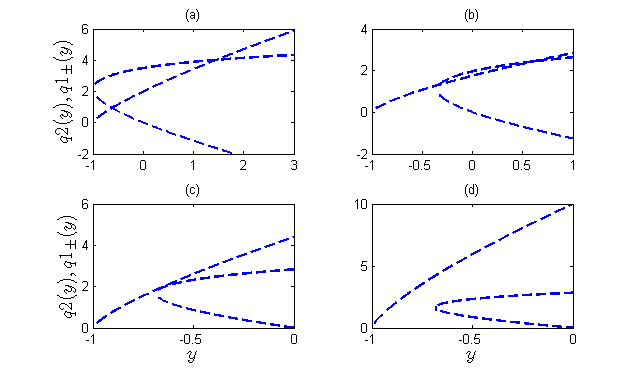}
\end{center}
\caption{
The four scenarios for $q_{1\pm}(y)$ and $q_2(y)$ crossings. Existence of a crossings correspond to ``Resonance-Phase" and no crossings correspond to ``No-Resonance Phase". (a) $\alpha=0.441$, $\gamma=2$ and $\omega=1$, for this case there will always be 2 crossings, $\alpha<\alpha_R$.   (b) $\alpha=0.6$, $\gamma=1.75$ and $\omega=1$, this panel corresponds to a ``Resonance-Phase" (c) $\alpha=0.5$, $\gamma=4.4$ and $\omega =1$, this panel describes a situation on the boundary between ``Resonance-Phase" and ``No-Resonance Phase". (d) $\alpha=0.5$, $\gamma=10$ and $\omega=1$, a ``No-Resonance Phase".
}
\label{fig12}
\end{figure}

We now argue that the boundary between ``Resonance-Phase" and ``No-Resonance Phase" is given by the following relation
\begin{equation}
\displaystyle{\frac{\omega}{\gamma^{\frac{1}{2-\alpha}}}=g(\alpha)},
\label{Sec2_16a}
\end{equation}
where $g(\alpha)$ is only $\alpha$ dependent and equals zero for $\alpha\leq \alpha_R$. One readily sees from Eq.~(\ref{Sec2_10}) that the large $y$ behavior of $q_{1\pm}(y)$ is proportional to $\pm\sqrt{y}$, where $q_2(y)$ behaves like $y^{1-\frac{\alpha}{2}}$ for large y (Eq.~(\ref{Sec2_11})), also we note that $\frac{dq_{1\pm}(y)}{dy}$ and $\frac{dq_2(y)}{dy}$ are monotonically decaying functions of $y$. As a result we have four options for the scenario of $q_2(y)$ crossing $q_{1\pm}(y)$ (see also Fig.~\ref{fig12}):\newline
\newline (i) $q_2(y_1)=q_{1-}(y1)$ and $q_2(y_2)=q_{1+}(y2)$ for $y1<y2$,
\newline (ii) $q_2(y1)=q_{1+}(y1)$ and $q_2(y2)=q_{1+}(y2)$ for $y1<y2$,
\newline (iii) $q_2(y)=q_{1+}(y)$ for a single $y$
\newline (iv) $q_2(y)\neq q_{1+}(y)$ and $q_2(y)\neq q_{1-}(y)$ for any $y$.
\newline\newline
When there are two crossings then the one with the larger $y$ corresponds to minimum and the smaller $y$ corresponds to maximum and belongs to the ``Resonance-Phase". When there are no crossings then we are in the ``No-Resonance Phase". The scenario (iii) corresponds exactly to the boundary between the two phases. In order to find the boundary, two conditions must be fulfilled
\begin{equation}
\displaystyle{q_2(y1)=q_{1+}(y1)}
\label{Sec2_17}
\end{equation}
and
\begin{equation}
\displaystyle{\frac{dq_2(y)}{dy}_{\mid y=y1} =
\frac{dq_{1+}(y)}{dy}_{\mid y=y1}},
\label{Sec2_18}
\end{equation}
as illustrated in panel (c) of Fig.~\ref{fig12}. Starting form Eq.~(\ref{Sec2_17}) we compare the left-hand side to some constant $c$ and using Eq.~(\ref{Sec2_11}) we find
\begin{equation}
y=c^{\frac{2}{2-\alpha}}\left(\frac{\omega}{\gamma^{\frac{1}{2-\alpha}}}\right)^2-1.
\label{Sec2_19}
\end{equation}
Comparing the right hand-side of Eq.~(\ref{Sec2_17}) to the same $c$ and using Eq.~(\ref{Sec2_19}) we arrive to the following relation
\begin{equation}
\displaystyle{\left(\frac{\omega}{\gamma^{\frac{1}{2-\alpha}}}\right)^2=
\frac{c^2-c\cos\left(\frac{\pi\alpha}{2}\right)(1+\frac{2}{\alpha})+\frac{2}{\alpha}}
{\frac{2}{\alpha}-c\cos\left(\frac{\pi\alpha}{2}\right)}c^{-\frac{2}{2-\alpha}}}.
\label{Sec2_20}
\end{equation}
Next performing the derivation in Eq.~(\ref{Sec2_18}) and using Eq.~(\ref{Sec2_19}) we find
\begin{equation}
\displaystyle{\left(\frac{\omega}{\gamma^{\frac{1}{2-\alpha}}}\right)^2=
\frac{(2-\alpha) c (2c-(1+\frac{2}{\alpha}\cos\left(\frac{\pi\alpha}{2}\right))}
{\frac{4}{\alpha}-4c\cos\left(\frac{\pi\alpha}{2}\right)
+\alpha c \cos\left(\frac{\pi\alpha}{2}\right)}c^{-\frac{2}{2-\alpha}}},
\label{Sec2_21}
\end{equation}
comparison of Eq.~(\ref{Sec2_21}) and Eq.~(\ref{Sec2_20}) supplies an equation for $c$
\begin{equation}
\begin{array}{l}
\alpha^3 \cos\left(\frac{\pi\alpha}{2}\right) c^3+
\left(2\alpha-5\alpha^2-\alpha(2+\alpha)\cos\left(\pi\alpha\right)\right)c^2+\\
\qquad\qquad\qquad
 12\alpha\cos\left(\frac{\pi\alpha}{2}\right) c-8=0.
\end{array}
\label{Sec2_22}
\end{equation}
Eq.~(\ref{Sec2_22}) has three different solutions where only one is real for $0<\alpha<1$, we will call it $c(\alpha)$ and
\begin{widetext}
\begin{equation}
\begin{array}{l}
\displaystyle{c(\alpha)=}
\\
\\
\displaystyle{
\frac{1}{3\alpha^3}\left[\sec\left(\frac{\pi\alpha}{2}\right)\left\{-2\alpha +5\alpha^2+2\alpha\cos\left(\pi\alpha\right) +\alpha^2\cos\left(\pi\alpha\right)\right.\right.}
\\

\displaystyle{
\left.\left.
-\left(2^{\frac{5}{3}}\alpha^{2}\sin^2\left(\frac{\pi\alpha}{2}\right)
\left(-4+20\alpha-7\alpha^2+(2+\alpha)^2\cos(\pi\alpha)\right)\right)
\right.\right.}
\\

\displaystyle{
\left.\left.
\left[-80\alpha^3+312\alpha^4-384\alpha^5+152\alpha^6-3\alpha^3\left(
-40+132\alpha-126\alpha^2+43\alpha^3\right)\cos(\pi\alpha)
-48\alpha^3\cos(2\pi\alpha)+72\alpha^4\cos(2\pi\alpha)
\right.\right.\right.}
\\

\displaystyle{
\left.\left.\left.
-24\alpha^6\cos(2\pi\alpha)+8\alpha^3\cos(3\pi\alpha)+12\alpha^4\cos(3\pi\alpha)
+6\alpha^5\cos(3\pi\alpha)+\alpha^6\cos(3\pi\alpha)
\right.\right.\right.}
\\

\displaystyle{
\left.\left.\left.
24\sqrt{6}\sqrt{-\alpha^9\cos^2\left(\frac{\pi\alpha}{2}\right)\left(
32-204\alpha+204\alpha^2-59\alpha^3+(2+\alpha)^2(-8+5\alpha)
\cos(\pi\alpha)\sin^4\left(\frac{\pi\alpha}{2}\right)\right)}\right]^{-\frac{1}{3}}
\right.\right.}
\\

\displaystyle{
\left.\left.
+\frac{1}{2^\frac{1}{3}}\left(
\left[-80\alpha^3+312\alpha^4-384\alpha^5+152\alpha^6-3\alpha^3\left(
-40+132\alpha-126\alpha^2+43\alpha^3\right)\cos(\pi\alpha)
-48\alpha^3\cos(2\pi\alpha)+72\alpha^4\cos(2\pi\alpha)
\right.\right.\right.\right.}
\\

\displaystyle{
\left.\left.\left.\left.
-24\alpha^6\cos(2\pi\alpha)+8\alpha^3\cos(3\pi\alpha)+12\alpha^4\cos(3\pi\alpha)
+6\alpha^5\cos(3\pi\alpha)+\alpha^6\cos(3\pi\alpha)
\right.\right.\right.\right.}
\\

\displaystyle{
\left.\left.\left.\left.
24\sqrt{6}\sqrt{-\alpha^9\cos^2\left(\frac{\pi\alpha}{2}\right)\left(
32-204\alpha+204\alpha^2-59\alpha^3+(2+\alpha)^2(-8+5\alpha)
\cos(\pi\alpha)\sin^4\left(\frac{\pi\alpha}{2}\right)\right)}\right]^{\frac{1}{3}}
\right)
\right]\right\}}.
\end{array}
\label{qwerf}
\end{equation}
\end{widetext}
We thus justified the use of Eq.~(\ref{Sec2_16a}) and $g(\alpha)$ is given by Eq.~(\ref{Sec2_23}).

\section{Appendix C: Exploration of ${d\chi ''(\Omega)}/{d\Omega}$}
\label{AppC}

We start with the exploration of ${d\chi ''(\Omega)}/{d\Omega}$, where $\chi ''(\Omega)$ is given by Eq.~(\ref{Sec3_01}),
\begin{widetext}
\begin{equation}
\begin{array}{l}
\displaystyle{
\frac{d\chi ''(\Omega)}{d\Omega}=
\frac{\gamma\Omega^{\alpha+3}\sin\left(\frac{\pi\alpha}{2}\right)}
{\left[(\omega^2-\Omega^2)^2+\gamma^2\Omega^{2\alpha}+2\gamma(\omega^2-\Omega^2)\Omega^{\alpha}
\cos\left(\frac{\pi\alpha}{2}\right)\right]^2}
\left[\alpha\left\{\left(\frac{w_z^2}{\Omega^2}-1\right)^2 + (\frac{\gamma}{\Omega^{2-\alpha}})^2 + \right.\right.
}\\
\displaystyle{
\left.\left.
2\left(\frac{\omega^2}{\Omega^2}-1\right)(\frac{\gamma}{\Omega^{2-\alpha}})
\cos\left(\frac{\pi\alpha}{2}\right)\right\}
-\left\{-4\left(\frac{\omega^2}{\Omega^2}-1\right) + 2\alpha(\frac{\gamma}{\Omega^{2-\alpha}})^2 +
2\left(\alpha\left(\frac{\omega^2}{\Omega^2}-1\right)-2\right)
(\frac{\gamma}{\Omega^{2-\alpha}})\cos\left(\frac{\pi\alpha}{2}\right)\right\}
\right],
}
\end{array}
\label{Sec3_02}
\end{equation}
\end{widetext}
and we easily see that in order to $\frac{d\chi ''(\Omega)}{d\Omega}=0$, the following condition must be fulfilled
\begin{equation}
\alpha\left(\frac{\gamma}{\Omega^{2-\alpha}}\right)^2 -
4\left(\frac{\gamma}{\Omega^{2-\alpha}}\right)\cos\left(\frac{\pi\alpha}{2}\right) -
(4y+\alpha y^2)=0,
\label{Sec3_03}
\end{equation}
where $y=\left(\frac{\omega^2}{\Omega^2}-1\right)>-1$.
The left hand side of Eq.~(\ref{Sec3_03}) is a second order polynomial in terms of $\frac{\gamma}{\Omega^{2-\alpha}}$, which is easily solved
\begin{equation}
\left(\frac{\gamma}{\Omega^{2-\alpha}}\right) =
\frac{4\cos\left(\frac{\pi\alpha}{2}\right)}{2\alpha} \pm
\frac{1}{2\alpha}\sqrt{16\cos^2\left(\frac{\pi\alpha}{2}\right) +
4\alpha(4y+\alpha y^2)}.
\label{Sec3_04}
\end{equation}
The right hand side of Eq.~(\ref{Sec3_04}) we will call $\tilde{q}_{1\pm}(y)$ and the left hand side $\tilde{q}_2(y)$,
\begin{equation}
\tilde{q}_2(y)=\frac{\gamma}{\omega^{2-\alpha}}(y+1)^{1-\frac{\alpha}{2}}.
\label{Sec3_05}
\end{equation}
The crossings of $\tilde{q}_{1\pm}(y)$ and $\tilde{q}_2(y)$ determine the extrema points of $\chi ''(\Omega)$ and using the fact that the for $y\rightarrow\infty$  $\tilde{q}_2(y)\propto y^{1-\frac{\alpha}{2}}$ and $\tilde{q}_{1\pm}(y)\propto y$  we have six different scenarios for the crossing of $\tilde{q}_2(y)$ and $\tilde{q}_{1\pm}(y)$.\newline
\newline (i)\qquad $\tilde{q}_2(y)=\tilde{q}_{1-}(y)$ for a single $y$ and there are no other crossings,
\newline (ii)\qquad $\tilde{q}_2(y_1)=\tilde{q}_{1-}(y_1)$ and $\tilde{q}_2(y_2)=\tilde{q}_{1+}(y_2)$ for $y_1<y_2$,
\newline (iii)\qquad  $\tilde{q}_2(y_1)=\tilde{q}_{1-}(y_1)$, $\tilde{q}_2(y_2)=\tilde{q}_{1+}(y_2)$ and $\tilde{q}_2(y_3)=\tilde{q}_{1+}(y_3)$ for $y_1<y_2<y_3$,
\newline (iv)\qquad $\tilde{q}_2(y_1)=\tilde{q}_{1+}(y_1)$, $\tilde{q}_2(y_2)=\tilde{q}_{1+}(y_2)$ and $\tilde{q}_2(y_3)=\tilde{q}_{1+}(y_3)$ for $y_1<y_2<y_3$,
\newline (v)\qquad $\tilde{q}_2(y_1)=\tilde{q}_{1+}(y_1)$ and $\tilde{q}_2(y_2)=\tilde{q}_{1+}(y_2)$ for $y_1<y_2$,
\newline (vi)\qquad $\tilde{q}_2(y)=\tilde{q}_{1+}(y)$ for a single $y$ and there are no other crossings.\newline\newline
Generally if there is only one crossing, scenario (i) and (vi), the meaning is that $\chi ''(\Omega)$ will have only one maximum and on the contrary when there are three crossings, scenario (iii) and (iv), there are two maximums and one minimum for $\chi ''(\Omega)$. These correspond to two different phases the ``One-Peak" phase and the ``Double-Peak" phase where the scenarios (ii) and (v) are the boundaries between these phases. We are interested in finding these boundaries, where for scenario (ii) and (v) two conditions must be fulfilled
\begin{equation}
\tilde{q}_2(y_1)=\tilde{q}_{1+}(y_1)
\label{Sec3_06}
\end{equation}
and
\begin{equation}
\displaystyle{\frac{d\tilde{q}_2(y)}{dy}_{\mid y=y_1} =
\frac{d\tilde{q}_{1+}(y)}{dy}_{\mid y=y_1}}.
\label{Sec3_07}
\end{equation}
Starting from Eq.~(\ref{Sec3_06}) we compare the left-hand side to some constant $\tilde{c}$ and using Eq.~(\ref{Sec3_05}) we find
\begin{equation}
y=\tilde{c}^{\frac{2}{2-\alpha}}\frac{\omega^2}{\gamma^{\frac{2}{2-\alpha}}}-1.
\label{Sec3_08}
\end{equation}
Comparing the right hand side of Eq.~(\ref{Sec3_06}) to the same $\tilde{c}$ and using Eq.~(\ref{Sec3_08}) we arrive to the following relation
\begin{equation}
\begin{array}{l}
\left(\frac{\omega}{\gamma^{\frac{1}{2-\alpha}}}\right)^2=\\
\left(
1-\frac{2}{\alpha}+\frac{1}{2}\sqrt{\frac{16}{\alpha^2} -
4\left(\frac{4\cos\left(\frac{\pi\alpha}{2}\right)}{\alpha}\tilde{c}-\tilde{c}^2
\right)}
\right)
\tilde{c}^{-\frac{2}{2-\alpha}}.
\end{array}
\label{Sec3_09}
\end{equation}
Next performing the derivation in Eq.~(\ref{Sec3_07}) and using Eq.~(\ref{Sec3_08}) we find
\begin{equation}
\begin{array}{l}
\left(\frac{\omega}{\gamma^{\frac{1}{2-\alpha}}}\right)^2=\\
\left(
\frac{1}{2}-\frac{1}{\alpha}+\frac{1}{2}\sqrt{\frac{(4-2\alpha)^2}{4\alpha^2}+
\tilde{c}\left(\frac{2-\alpha}{\alpha}\right)(2\alpha\tilde{c} -
4\cos\left(\frac{\pi\alpha}{2}\right))
}
\right)
\tilde{c}^{-\frac{2}{2-\alpha}},
\end{array}
\label{Sec3_10}
\end{equation}
comparison of Eq.~(\ref{Sec3_09}) and Eq.~(\ref{Sec3_10}) supplies an equation for $\tilde{c}$
\begin{equation}
\begin{array}{l}
4\alpha^2\tilde{c}^4 -
16\cos\left(\frac{\pi\alpha}{2}\right)(2+\alpha)\tilde{c}^3
\\
+
\frac{16}{\alpha^2}\left( -4 +8\alpha-\alpha^2 +
\cos^2\left(\frac{\pi\alpha}{2}\right)(2+\alpha)^2\right)\tilde{c}^2
\\
 -
\frac{64}{\alpha^2}\cos\left(\frac{\pi\alpha}{2}\right)(6-\alpha)\tilde{c} +
\frac{64}{\alpha^3}(4-\alpha)=0.
\end{array}
\label{Sec3_11}
\end{equation}
Eq.~(\ref{Sec3_11}) is a forth order polynomial and could pe solved
by standard methods or using Mathematica. It has $4$ different
solutions while two of the solutions have non-zero Imaginary parts
for any $0<\alpha<1$, while the other two solutions have no
Imaginary part for $\alpha<\alpha_{\chi_2}\approx 0.70776$. Lets
call these solutions $\tilde{c}_1(\alpha)$ and
$\tilde{c}_2(\alpha)$, the non-zero Imaginary part for
$\alpha>\alpha_{\chi_2}$ of both $\tilde{c}_1(\alpha)$ and
$\tilde{c}_2(\alpha)$ means that only scenario (vi) is applicable
for such $\alpha$'s and we are in the ``One-Peak" phase. The
boundaries between the two phases are given by
\begin{equation}
\begin{array}{l}
\frac{\omega}{\gamma^{\frac{1}{2-\alpha}}}=
\tilde{g}_{1,2}(\alpha)=
\\
\sqrt{
1-\frac{2}{\alpha}+\frac{1}{2}\sqrt{\frac{16}{\alpha^2} -
4\left(\frac{4\cos\left(\frac{\pi\alpha}{2}\right)}{\alpha}\tilde{c}_{1,2}(\alpha)
-\tilde{c}_{1,2}^2(\alpha)
\right)}
}
\tilde{c}_{1,2}^{-\frac{1}{2-\alpha}}(\alpha)
\end{array}
\label{Sec3_12}
\end{equation}
where the subscript $1$ is for $\tilde{g}_1(\alpha)$ the lower bound
in Fig.~\ref{fig14} and subscript $2$ is the upper bound
$\tilde{g}_2(\alpha)$ in Fig.~\ref{fig14}. For $\tilde{g}_1(\alpha)$
there is also another interesting point $\alpha_{\chi_1}=0.527031$
which satisfies the following relation
\begin{equation}
\alpha_{\chi_1}^2-4\alpha_{\chi
1}+4\cos^2\left(\frac{\pi\alpha_{\chi_1}}{2}\right)=0,
\label{Sec3_13}
\end{equation}
for $\alpha<\alpha_{\chi_1}$ $\tilde{g}_1(\alpha)=0$.


\begin{thebibliography}{999}

\bibitem{Adelman} S. A. Adelman {J. Chem. Phys.} {\bf64}, 124 (1976).

\bibitem{Lutz} E. Lutz, {Phys. Rev. E.} {\bf64}, 051106 (2001).

\bibitem{Pottier} N. Pottier, {Physica A} {\bf317}, 371 (2003).

\bibitem{Chaud} S. Chaudhury and B. J. Cherayil {J. Chem. Phys.} {\bf125}, 024904 (2006).

\bibitem{Hangii} J. D. Bao, P. H\"{a}nggi, and Y. Z. Zhuo {Phys. Rev. E.} {\bf72}, 061107 (2005).



\bibitem{Kop} R. Kopferman {J. of Statistical Physics} {\bf114}, 291 (2004).

\bibitem{Goychuk3} I. Goychuk and P. H\"{a}ngii, {Phys. Rev. Lett.} {\bf 99}, 200601 (2007).

\bibitem{Metzler1} R. Metzler and J. Klafter {Phys. Rep.} {\bf339}, 1 (2001).

\bibitem{Xie3} H. Yang, G. Luo, P. Karnchanaphanurach, T. Louie, I. Rech, S. Cova, L. Xun, and X.S. Xie1 {Science} {\bf302}, 262, (2003).

\bibitem{Xie1} W. Min, G. Lou, B.J. Cherayil, S.C. Kou, and X.S. Xie {Phys. Rev. Lett.} {\bf94},
198302 (2005).

\bibitem{Kibble} T. W. Kibble and F. H. Berkshire {\em Classical Mechanics } {Longman, London, 1996}.

\bibitem{Argentina} A. D. Vi\~{n}ales and M. A. Desp\'{o}sito {Phys. Rev. E.} {\bf73}, 016111 (2006).

\bibitem{Burov} S. Burov and E. Barkai, {Phys. Rev. Lett.} {\bf 100}, 070601 (2008).

\bibitem{Mandelb} B. Mandelbrot and J. Van Ness, {SIAM Rev.} {\bf10}, 422 (1968)
\bibitem{Xie2} S. C. Kou and X. Sunney Xie, {Phys. Rev. Lett.} {\bf93}, 180603 {2004}.

\bibitem{Kubo1} R. Kubo, M. Toda, and N. Hashitsume, {\em Statistical Physics II, Nonequilibrium Statistical Mechanics} (Springer-Verlag, Berlin, 1985).

\bibitem{Samko} S. G. Samko, A. A. Kilbas, and O. I. Marichev, {\em Fractional Integrals and
Derivatives and Their applications} (Nauka i Technika, Minsk, 1987) (in Russian).

\bibitem{Miller} K. S. Miller and B. Ross, {\em An Introduction to the Fractional Differential
Equations} (Wiley, New York, 1993).

\bibitem{Masoliver1} J. M. Porr\`{a}, K.G. Wang, and J. Masoliver, {Phys. Rev. E} {\bf53}, 5872 (1996).

\bibitem{Zwanzig} R. Zwanzig, {J. Stat. Phys.} {\bf9}, 215 (1973).

\bibitem{Barkai2} E. Barkai and R. J. Silbey {J. Phys. Chem.} {\bf104}, 3866 (2000).

\bibitem{Sokol} I. M. Sokolov, J. Klafter, and A. Blumen, {Phys. Today} {\bf55}, No. 11, 48 (2002).

\bibitem{Barkai1} E. Barkai, {Phys. Rev. E} {\bf63}, 046118 (2001).

\bibitem{Barkai3} R. Metzler, E. Barkai, and J. Klafter, {Phys. Rev. Lett.} {\bf82}, 3563 (1999) .

\bibitem{Achar} B. N. Narahari Achar, J. W. Hanneken and T. Enck, T. Clarke {Physica A} {\bf297}, 361 (2001).

\bibitem{Ryabov} Ya. E. Ryabov and A. Puzenko {Phys. Rev. B.} {\bf66}, 184201 (2002).

\bibitem{Zaslav} G. M. Zaslavsky, A. A. Stanislavsky, and M. Edelman, {Chaos} {\bf16} 013102 (2006).

\bibitem{Kilbas} A. A. Kilbas, H. M. Srivastava, and J. J. Trujillo {\em Theory and applications of fractional differential equations} {Elsevier, Amstedam, 2006}.

\bibitem{Adding} According to Ref.~\cite{Kilbas}, such a fractional oscillator was considered by F. Mainardi in {W. F. Ames (Ed.), {}\em 12th IMACS World Congress,} vol. 1, {Atlanta, 1994} 329-333.

\bibitem{Lavr} M. A. Lavrentiev and B. V. Shabat, {\em Methods of the Theory of Functions of Complex Variable}
(Nauka, Moscow, 1973).


\bibitem{Doech} G. Doetsch {\em Guide to the applications of Laplace transforms} {Van. Nostrand, London, 1961}.

\bibitem{Abram} M. Abramowitz and I. Stegun, {\em Handbook of Mathematical Functions with Formulas,
Graphs, and Mathematical Tables} (Dover, New York, 1971).


\bibitem{Erdelyi} A. Erd\`{e}lyi {\em Tables of Integral Transforms} (McGraw-Hill, New York, 1954).

\bibitem{Feller} W. Feller, {\em An introduction to probability theory and
its applications, Volume 2} (John Wiley and Sons, New York, 1971).


\bibitem{Barbi} F. Barbi, M. Bologna, and P. Grigolini, \emph{Phys. Rev. Lett.} {\bf95}, 220601 (2005).

\bibitem{Sokolov2} I. M. Sokolov, and J. Klafter, \emph{Phys. Rev. Phys.} {\bf97}, 140602 (2006).

\bibitem{Goychuk1} E. Heinsalu, M. Patriarca, I. Goychuk, and P. H\"{a}nggi, \emph{Phys. Rev. Lett.} {\bf99}, 120602 (2007).

\bibitem{Weron} K. Weron and M. Kotulski, {Physica A} {\bf 232}, 180 (1996).

\bibitem{Goychuk2} I. Goychuk, \emph{Phys. Rev. E.} {\bf76} 040102(R) (2007).

\bibitem{Granek} R. Granek and J. Klafter {Phys. Rev. Lett.} {\bf95}, 098106 (2005).


\bibitem{Luo} G. Luo, I. Andricionoaei, X. S. Xie, and M. Karplus {J. Phys. Chem. B} {\bf110}, 9363 (2006).

\bibitem{Coffey1} W. T. Coffey, Yu. P. Kalmykov, and S. V. Titov {Phys. Rev. E.} {\bf65}, 032102 (2002).

\bibitem{Coffey2} W. T. Coffey, Yu. P. Kalmykov, and S. V. Titov {Phys. Rev. E.} {\bf65} 051105 (2002).

\bibitem{Coffey3} W. T. Coffey, Yu. P. Kalmykov, and J. T. Waldron, {\em The Langevin equation : with applications to stochastic problems in physics, chemistry, and electrical engineering} (World Scientific, New Jersey, 2004).

\bibitem{Rocard} M. Y. Rocard {J. Phys. Radium} {\bf4}, 247 (1933).

\bibitem{Scaife} B. K. P. Scaife, {\em Principles of Dielectrics} (Oxford University Press, London, 1989).



\bibitem{Dachwitz} E. Dachwitz, F. Parak, and M. Stockhausen, {Ber. Bunsenges. Phys. Chem.} {\bf93}, 1454 (1989).

\bibitem{Gotze} W. G\"{o}tze and L. Sj\"{o}rgen, {Rep. Prog. Phys.} {\bf55}, 241 (1992).

\bibitem{Bagchi} N. Nandi, and B. Bagchi, {J. Phys. Chem. A} {\bf102}, 8217 (1998).


\end{thebibliography}
\end{document}